\documentclass[utf8]{frontiersSCNS} 

\usepackage{url,hyperref,microtype,subcaption}
\usepackage[onehalfspacing]{setspace}
\usepackage{booktabs}
\usepackage{array}
\usepackage{setspace}
\usepackage{mathtools}
\usepackage{graphicx}
\usepackage[draft,inline,nomargin]{fixme}
\usepackage{color}

\usepackage{multirow}

\newcommand{\rme}{\mathrm{e}}

\newcommand{\eref}[1]{Equation~(\ref{#1})} 
  
\newcommand{\Fref}[1]{Figure~\ref{#1}}  \newcommand{\Tref}[1]{Table~\ref{#1}}

\usepackage{pdfpages}

\def\keyFont{\fontsize{8}{11}\helveticabold }
\def\firstAuthorLast{Morales {et~al.}} 
\def\Authors{Jos{\'e} A. Morales\,$^{1,2}$, Ewan Colman\,$^{3,4}$, Sergio S\'anchez\,$^{2}$, Fernanda S\'anchez-Puig\,$^{1}$, Carlos Pineda\,$^{2,5}$, Gerardo I{\~n}iguez\,$^{6,7}$, Germinal Cocho\,$^{2}$, Jorge Flores\,$^{2}$, and Carlos Gershenson\,$^{3,4,8,9,*}$}
\def\Address{$^{1}$Facultad de Ciencias, Universidad Nacional Aut{\'o}noma de M{\'e}xico, 01000 M\'exico D.F., Mexico\\
$^{2}$ Instituto de F{\'i}sica, Universidad Nacional Aut{\'o}noma de M{\'e}xico, 01000 M\'exico D.F., Mexico\\
$^{3}$ Instituto de Investigaciones en Matem{\'a}ticas Aplicadas y en Sistemas, Universidad Nacional Aut{\'o}noma de M{\'e}xico, 01000 M\'exico D.F., Mexico\\
$^{4}$ Centro de Ciencias de la Complejidad, Universidad Nacional Aut{\'o}noma de M{\'e}xico, 01000 M\'exico D.F., Mexico\\
$^{5}$ Faculty of Physics, University of Vienna, 1090 Wien, Austria\\
$^{6}$ Next Games, 00530 Helsinki, Finland\\
$^{7}$ Department of Computer Science, Aalto University School of Science, 00076 Aalto, Finland\\
$^{8}$ SENSEable City Lab, Massachusetts Institute of Technology, 02139 Cambridge, MA, USA\\
$^{9}$ ITMO University, 199034 St. Petersburg, Russian Federation}
\def\corrAuthor{Carlos Gershenson}
\def\corrEmail{cgg@unam.mx}

\begin{document}
\onecolumn
\firstpage{1}

\title[Rank Dynamics of Word Usage]{
Rank Dynamics of Word Usage\\ at Multiple Scales} 

\author[\firstAuthorLast ]{\Authors} 
\address{} 
\correspondance{} 

\extraAuth{}

\maketitle

\begin{abstract}

\section{}
The recent dramatic increase in online data availability has allowed researchers to explore human culture with unprecedented detail, such as the growth and diversification of language. In particular, it provides statistical tools to explore whether word use is similar across languages, and if so, whether these generic features appear at different scales of language structure. Here we use the Google Books $N$-grams dataset to analyze the temporal evolution of word usage in several languages. We apply measures proposed recently to study rank dynamics, such as the diversity of $N$-grams in a given rank, the probability that an $N$-gram changes rank between successive time intervals, the rank entropy, and the rank complexity. Using different methods, results show that there are generic properties for different languages at different scales, such as a core of words necessary to minimally understand a language. We also propose a null model to explore the relevance of linguistic structure across multiple scales, concluding that $N$-gram statistics cannot be reduced to word statistics. We expect our results to be useful in improving text prediction algorithms, as well as in shedding light on the large-scale features of language use, beyond linguistic and cultural differences across human populations.

\tiny
 \keyFont{ \section{Keywords:} Culturomics, N-grams, language evolution, rank diversity, complexity} 
\end{abstract}
\section{Introduction} 

The recent availability of large datasets on language, music, and other cultural constructs
has allowed the study of human culture at a level never possible before, opening
the data-driven field of {\it culturomics}~\citep{Lieberman2007,Michel14012011, dodds2011temporal, serra2012measuring, blumm2012dynamics, CPLX:CPLX21436, e15125084,PhysRevX.3.021006,perc2013self, Febres2013Complexity-meas, Wagner2014, Pina2016, Pina-Garcia2018}. In the social sciences
and humanities, lack of data has traditionally made it difficult or even impossible to
contrast and falsify theories of social behaviour and cultural evolution. Fortunately, digitalized data and computational algorithms allow us to tackle
these problems with a stronger statistical
basis~\citep{Wilkens2015}. In particular, the Google Books
$N$-grams dataset~\citep{Michel14012011, 
wijaya2011understanding, petersen2012languages, 
petersen2012statistical, Perc07122012,
10.1371/journal.pone.0059030, 
Ghanbarnejad20141044, Dodds24022015, PhysRevX.6.021009} continues to be a fertile source of analysis in culturomics, since it contains an estimated $4\%$ of all books printed throughout the world until 2009. From the 2012 update of this public dataset, we measure frequencies per year of words (1-grams),
pairs of words (2-grams), up until $N$-grams with $N = 5$ for several languages, and focus on how scale (as measured by $N$) determines the statistical and temporal characteristics of language structure.

We have previously studied the temporal evolution of
word usage (1-grams) for six Indo-European languages: English, Spanish,
French, Russian, German, and Italian, between 1800 and 2009~\citep{Cocho2015}. We first analysed the language rank distribution~\citep{zipf,newman2005power,1367-2630-13-4-043004,PhysRevE.83.036115}, \emph{i.e.} the set of all words
ordered according to their usage frequency. 
By making fits of this rank distribution with several models, we noticed
that no single functional shape fits all languages well. Yet, we also found regularities on how ranks of words change in time: Every year, the most frequent
word in English (rank 1) is `\emph{the}',
while the second most frequent word (rank 2) is `\emph{of}'.
However, as the rank $k$ increases, the number of words occupying the $k$-th place of usage (at some point in time) also increases. Intriguingly, we observe the same generic behaviour in the temporal evolution of performance rankings in some sports and games~\citep{Morales2016}.

To characterize this generic feature of rank dynamics, we have proposed
the {\it rank diversity} $d(k)$ as the number of words occupying a given
rank $k$ across all times, divided by the number $T$ of time intervals considered (for \citet{Cocho2015}, $T=210$ intervals of one year). For example, in English $d(1)=1/210$,
as there is only one word (`\emph{the}')
occupying $k=1$ every year. The rank diversity increases with
$k$, reaching a maximum $d(k)=1$ when there is a different word
at rank $k$ each year. The rank diversity curves of all six languages
studied can be well approximated by a sigmoid curve, suggesting that
$d(k)$ may reflect generic properties of language evolution, irrespective of differences in grammatical structure and cultural features of language use.
Moreover, we have found rank diversity useful to estimate
the size of the {\it core} of a language, \emph{i.e.}
the minimum set of words necessary to speak and understand a tongue~\citep{Cocho2015}.

In this work, we extend our previous analysis of rank dynamics to $N$-grams with $N =
1, 2, ... 5$ between 1855 and 2009 ($T = 155$) for the same six languages, considering the first $11,140$ ranks in all 30 datasets (to have equal size and avoid potential finite-size effects). In the next section, we present results for the rank diversity of $N$-grams. We then compare empirical digram data with a null expectation for 2-grams that are randomly generated from the monogram frequency distribution. Results for novel measures of change probability, rank entropy, and rank complexity follow. Next, we discuss the implications of our results, from practical applications in text prediction algorithms, to the emergence of generic, large-scale features of language use despite the linguistic and cultural differences involved. Details of the methods used close the paper.

\section{Results} 

\subsection{Rank Diversity of $N$-gram usage}

\begin{figure}[h] 
\begin{center}
\includegraphics{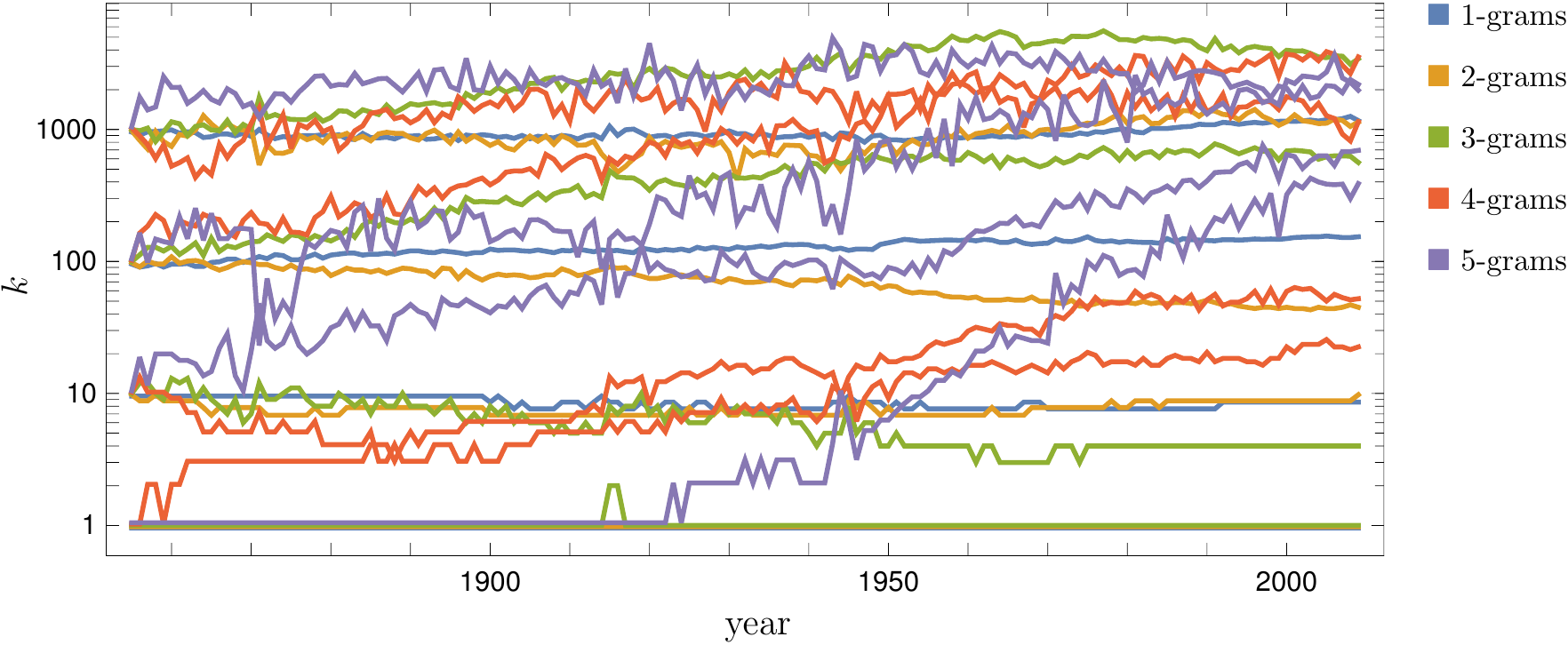}
\end{center}
\caption{{\bf Rank evolution of $N$-grams in French}. Rank trajectories across
time for $N$-grams ($N = 1, \ldots, 5$) that are initially in rank $1$, $10$,
$100$ and $1000$ at the year 1855. The plot is semilogarithmic, so similar
changes across ranks correspond to changes proportional to the rank itself.
Other languages (not shown) behave in a similar way: changes are more frequent as $N$ increases. We have added a small shift over the $y$-axis for some curves, to see more clearly how the most frequently used $N$-grams remain at $k = 1$ for long periods of time.}
\label{fig:espaguetis}
\end{figure}

\begin{figure} 
\begin{center}
\includegraphics{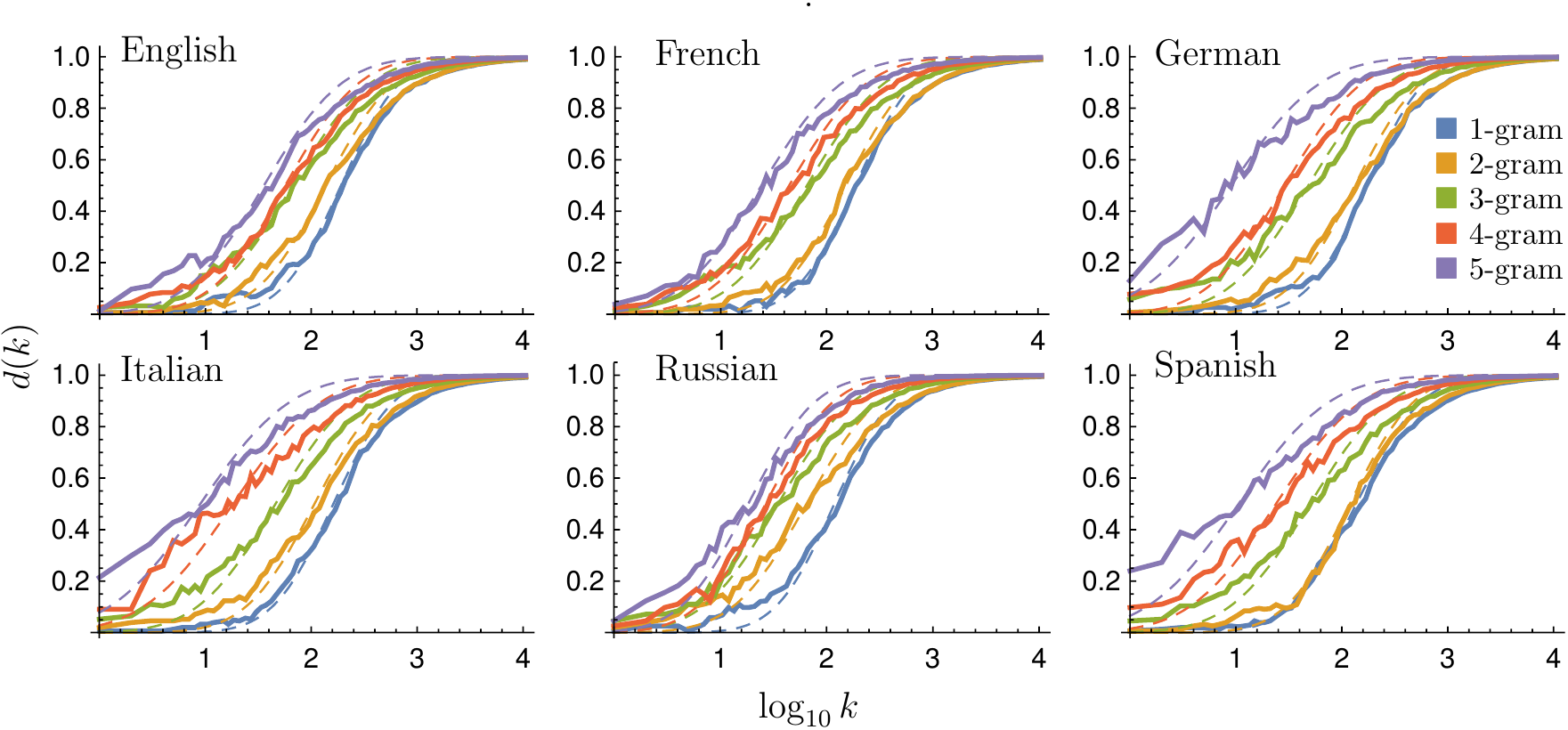}
\end{center}
\caption{
{\bf Rank diversity for different languages and $N$-grams}. Binned rank diversity $d(k)$ as a
function of rank $k$ for all languages and $N$ values considered (continuous
lines). We also include fits according to the 
sigmoid in \eref{eq:sigmoid} (dashed lines), with $\mu$, $\sigma$ and the associated $e$ error summarised in Table~\ref{tab:rankdiv_params}. Windowing is
done averaging $d(k)$ every 0.05 in $\log_{10} k$.
}
\label{fig:d}
\end{figure}

Figure~\ref{fig:espaguetis} shows the rank trajectories across time for selected $N$-grams in French, classified by value of $N$ and their rank of usage in the first year of measurement (1855). The behaviour of these curves is similar for all languages: $N$-grams in low ranks (most frequently used) change their position less than $N$-grams in higher ranks, yielding a sigmoid rank diversity $d(k)$ (Figure~\ref{fig:d}). Moreover, as $N$ grows, the rank diversity tends to be larger, implying a larger variability in the use of particular phrases relative to words. To better grasp how $N$-gram usage varies in time, Tables S1-S30 in the Supplementary Information list the top $N$-grams in several years for all languages.
We observe that the lowest ranked $N$-grams (most frequent)
tend to be or contain function words (articles, prepositions, conjunctions), since their use is largely independent of the text topic.
On the other hand, content words (nouns, verbs, adjectives, adverbs) are contextual,
so their usage frequency varies widely across time and texts. Thus, we find it reasonable that top $N$-grams vary more in time for larger $N$.

As Figure~\ref{fig:d} shows, rank diversity $d(k)$ tends to grow with the scale $N$ since, as $N$ increases, it is less probable
to find $N$-grams with only function words (especially in Russian, which
has no articles). For $N=1, 2$ in some languages, function words dominate the top ranks, decreasing their diversity, while the most popular content words (1-grams) change rank
widely across centuries. Thus, we expect
the most frequent 5-grams to change relatively more in time (for example, in Spanish, $d(1)$ is $\frac{1}{155}$ for 1-grams and 2-grams, $\frac{7}{155}$ for 3-grams, $\frac{15}{155}$ for 4-grams, and finally $\frac{37}{155}$ for 5-grams). Overall, we observe that all rank diversity curves can be well fitted
by the sigmoid curve
\begin{equation}
\Phi_{\mu,\sigma}(\log_{10} k)=\frac{1}{\sigma\sqrt{2\pi}}   \int_{-\infty}^{\log_{10} k} \rme^{-\frac{(y-\mu)^2}{2\sigma^2}} {\rm d}y, 
\label{eq:sigmoid}
\end{equation}
where $\mu$ is the mean and $\sigma$ the standard deviation of the sigmoid, both dependent on language and $N$ value (\Tref{tab:rankdiv_params}). 

\begin{figure} 
\begin{center}
\includegraphics{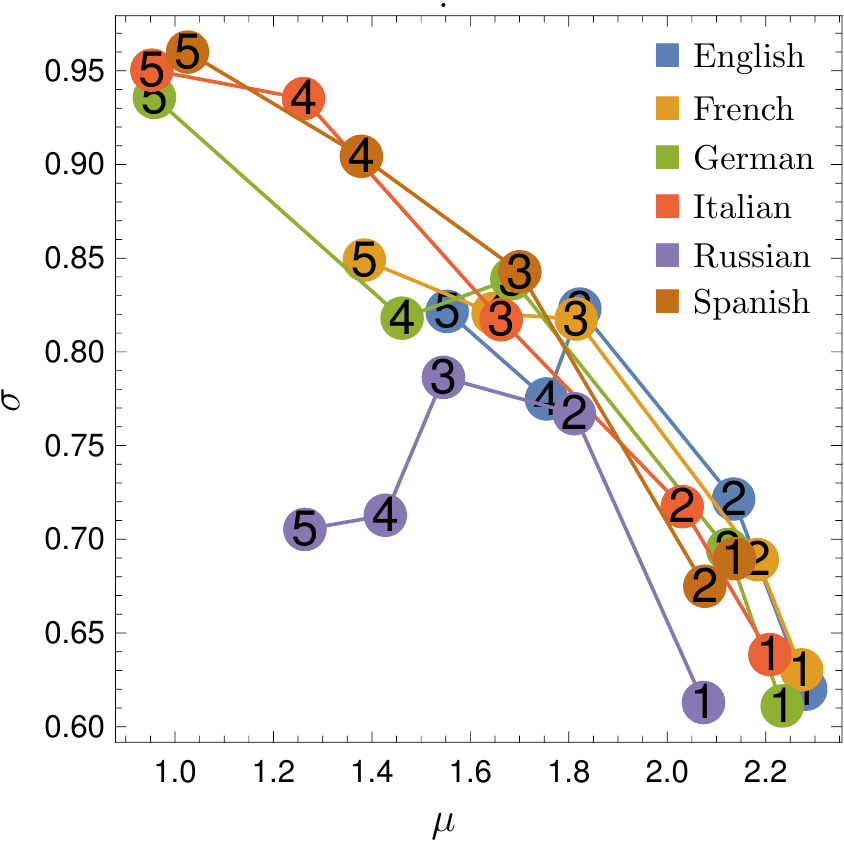}
\end{center}
\caption{{\bf Fitted parameters for rank diversity}. Parameters $\mu$ and $\sigma$ for the sigmoid fit of the rank diversity $d(k)$, for all languages (indicated by colors) and $N$ values (indicated by numbers). We observe an (approximate) inversely proportional relation between $\mu$ and $\sigma$.}
\label{fig:clusters}
\end{figure}

In Figure~\ref{fig:clusters} we see the fitted values of $\mu$ and $\sigma$ for all datasets considered. In all cases $\mu$ decreases with $N$, while in most cases $\sigma$ increases with $N$, roughly implying an inversely proportional relation between $\mu$ and $\sigma$. 
It is interesting to note that for Romance languages (Spanish, French, and Italian), $\sigma$ increases when moving from 3-grams to 5-grams, while for Germanic languages (English and German) and Russian (a Slavic language), there is a decrease in $\sigma$ from $N=3$ to $N=4$.

\subsection{Null model: random shuffling of monograms}

In order to understand the dependence of language use --- as measured by $d(k)$ --- on scale ($N$), we can ask whether the statistical properties of $N$-grams can be deduced exclusively from those of monograms, or if the use of higher-order $N$-grams reflects features of grammatical structure and cultural evolution that are not captured by word usage frequencies alone. 
To approach this question, we consider a null model of language in which grammatical structure does not influence the order of words. We base our model on the idea of shuffling 1-gram usage data to eliminate the grammatical structure of the language, while preserving the frequency of individual words (more details in Methods, Section \ref{null}).

\subsubsection{Rank diversity in null model} 

\begin{figure} 
\begin{center}
\includegraphics[width=\textwidth]{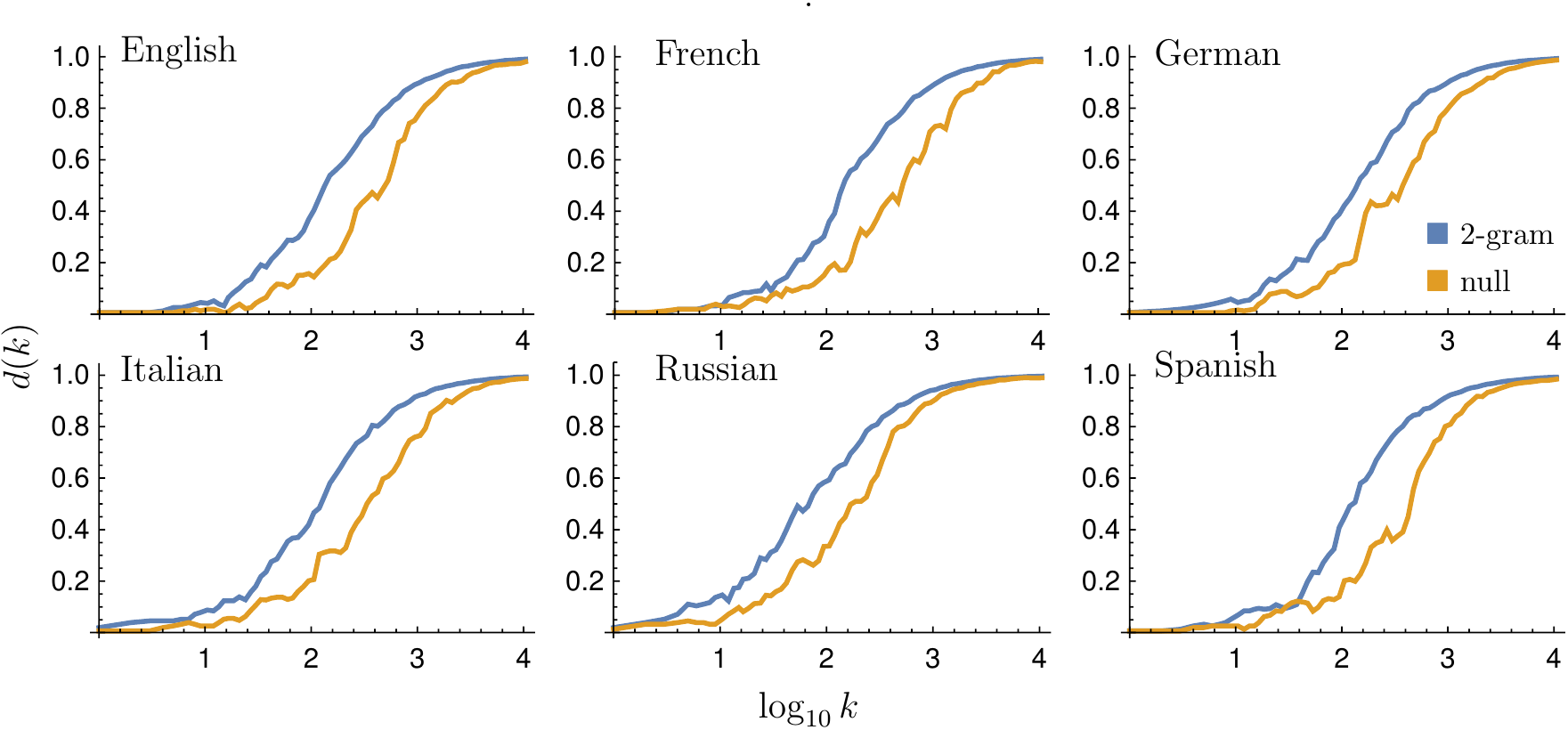}
\end{center}
\caption{{\bf Rank diversity in the null model}. Rank diversity $d(k)$ of both empirical and randomly generated 2-grams. The rank diversity of the null model tends to be to the right of that of the data, \emph{i.e.} lower. Windowing in this and similar figures is done in the same way as in Figure~\ref{fig:d}.}
\label{fig:shuffled}
\end{figure}

As can be seen in Figure ~\ref{fig:shuffled}, the rank diversity of digrams constructed from shuffled monograms is generally lower than for the non-shuffled digrams, although it keeps the same functional shape of \eref{eq:sigmoid} (see fit parameters in \Tref{tab:rankdiv_params}). In the absence of grammatical structure, the frequency of each 2-gram is determined by the frequencies of its two constituent 1-grams. Thus, combinations of high frequency 1-grams dominate the low ranks, including some that are not grammatically valid --- \emph{e.g.} {\it 'the the'}, {\it 'the of'}, {\it 'of of'} --- but are much more likely to occur than most others.  Moreover, the rank diversity of such combinations is lower than we see in the non-shuffled data because the low ranked 1-grams that create these combinations are relatively stable over time. Thus, we can conclude that the statistics of higher order $N$-grams is determined by more than word statistics, \emph{i.e.} language structure matters at different scales.

\subsubsection{$z$-scores in null model} 

The amount of structure each language exhibits can be quantified by the $z$-scores of the empirical 2-grams with respect to the shuffled data. Following its standard definition, the $z$-score of a 2-gram is a measure of the deviation between its observed frequency in empirical data and the frequency we expect to see in a shuffled dataset, normalized by the standard deviation seen if we were to shuffle the data and measure the frequency of the 2-gram many times (see Section \ref{null} for details).

\begin{figure} 
\centering
\includegraphics[width=\textwidth]{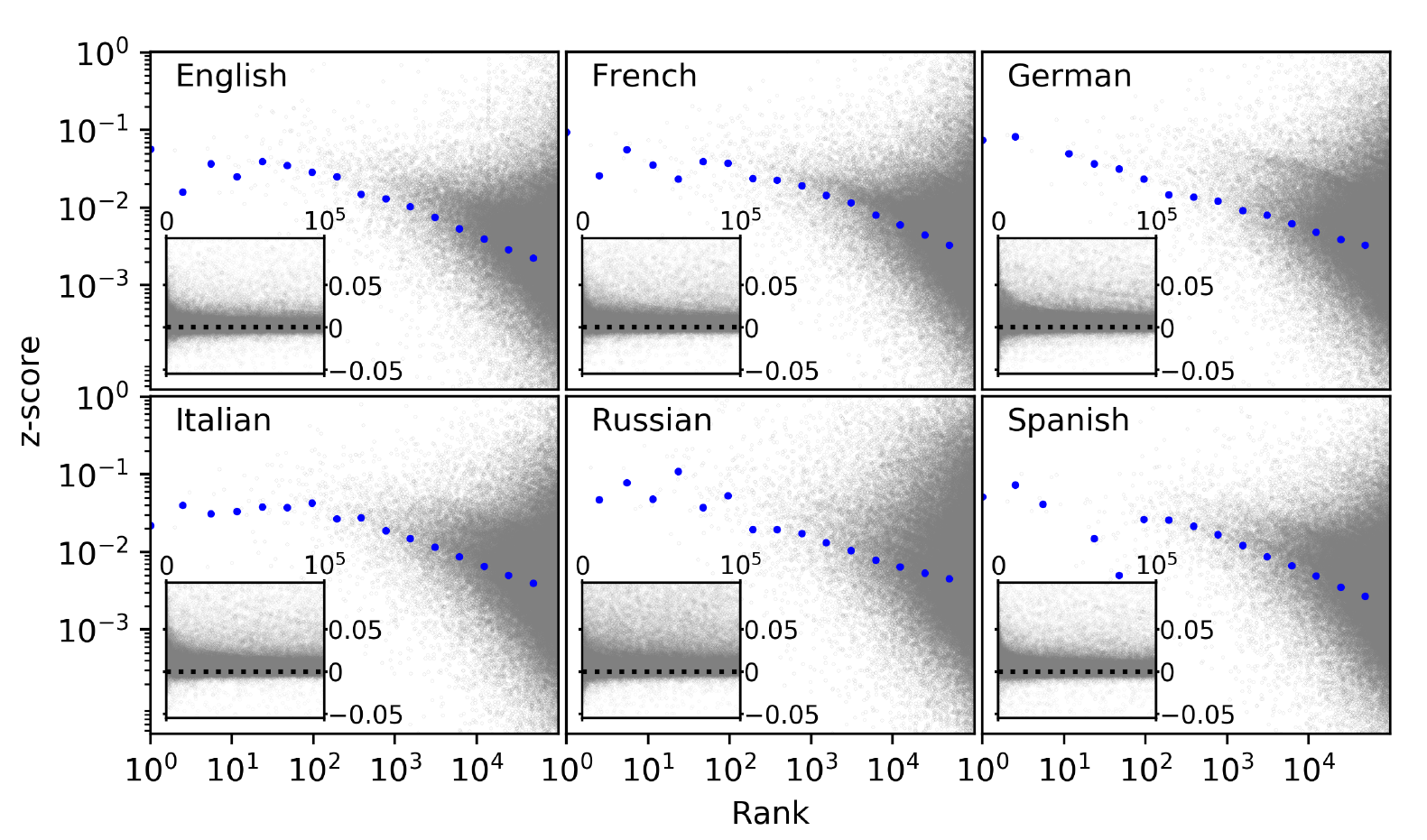}
\caption{{\bf $z$-scores between empirical and null model digrams}. $z$-scores, calculated from Eq.\eqref{normalized_z}, for the top $10^{5}$ 2-grams in $2008$. Each point represents a 2-gram found in the empirical data. The blue dots show the median $z$-score of logarithmically binned values (first bin contains only the first rank; each consecutive bin is twice as large as the previous one). The inset shows the same values on a linear scale $y$ axis, with a dashed line indicating $z=0$.}
\label{z_scores}
\end{figure}

The 2-grams with the highest $z$-scores are those for which usage of the 2-gram accounts for a large proportion of the usage of each of its two constituent words. That is, both words are more likely to appear together than they are in other contexts (for example, {\it `led zeppelin'} in the Spanish datasets), suggesting that the combination of words may form a linguistic token that is used in a similar way to an individual word. We observe that the majority of 2-grams have positive $z$-scores, which simply reflects the existence of non-random structure in language (Figure \ref{z_scores}). What is more remarkable is that many 2-grams, including some at low ranks ({\it `und der'}, {\it `and the'}, {\it `e di'}), have negative $z$-scores; a consequence of the high frequency and versatility of some individual words.

After normalizing the results to account for varying total word frequencies between different language datasets, we see that all languages exhibit a similar tendency for the $z$-score to be smaller at higher ranks (measured by the median; this is not the case for the mean). This downward slope can be explained by the large number of 2-grams that are a combination of one highly versatile word, \emph{i.e.} one that may be combined with a diverse range of other words, with relatively low frequency words (for example {\it 'the antelope'}). In such cases, $z$-scores decrease with rank as $z\sim k^{-1/2}$ (see Section \ref{null}).

\subsection{Next-word entropy} 

\begin{figure} 
\centering
\includegraphics[width=0.5\textwidth]{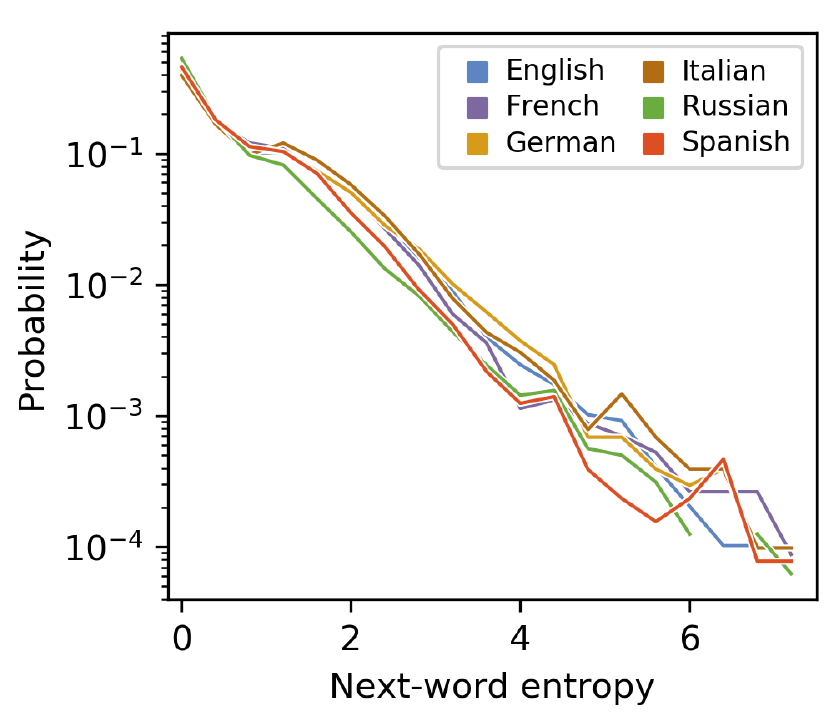}
\caption{{\bf Next-word entropy for different languages and null model}. Probability distribution of next-word entropies, calculated using Eq.~\eqref{eq:Enw}. The range of entropies is segregated into bins of width $1/2$, while the probability is calculated as the number of words whose next-word entropy falls inside the bin, divided by the total number of words.}
\label{nwe_fig}
\end{figure}

Motivated by the observation that some words appear alongside a diverse range of other words, whereas others appear more consistently with the same small set of words, we examine the distribution of next-word entropies. Specifically, we define the {\it next-word entropy} for a given word $i$ as the (non-normalized) Shannon entropy of the set of words that appear as the second word in 2-grams for which $i$ is the first. In short, the next-word entropy of a given word quantifies the difficulty of predicting the following word. As shown in Figure \ref{nwe_fig}, words with higher next-word entropy are less abundant than those with lower next-word entropy, and the relationship is approximately exponential. 

\subsection{Change probability of $N$-gram usage}

To complement the analysis of rank diversity, we propose a related measure: the change probability $p(k)$, \emph{i.e.} the probability that a word at rank $k$ will change rank in one time interval. We calculate it for a given language dataset by dividing the number of times elements change for given rank $k$ by the number of temporal transitions, $T - 1$. The change probability behaves similarly to rank diversity in some cases. For example, if there are only two $N$-grams that appear with rank 1, $d(1)=2/155$. If one word was ranked first until 1900 and then a different word became first, there was only one rank change, thus $p(1)=1/154$. However, if the words alternated ranks every year (which does not occur in the datasets studied), the rank diversity would be the same, but $p(1)=1$. 

\begin{figure} 
\begin{center}
\includegraphics{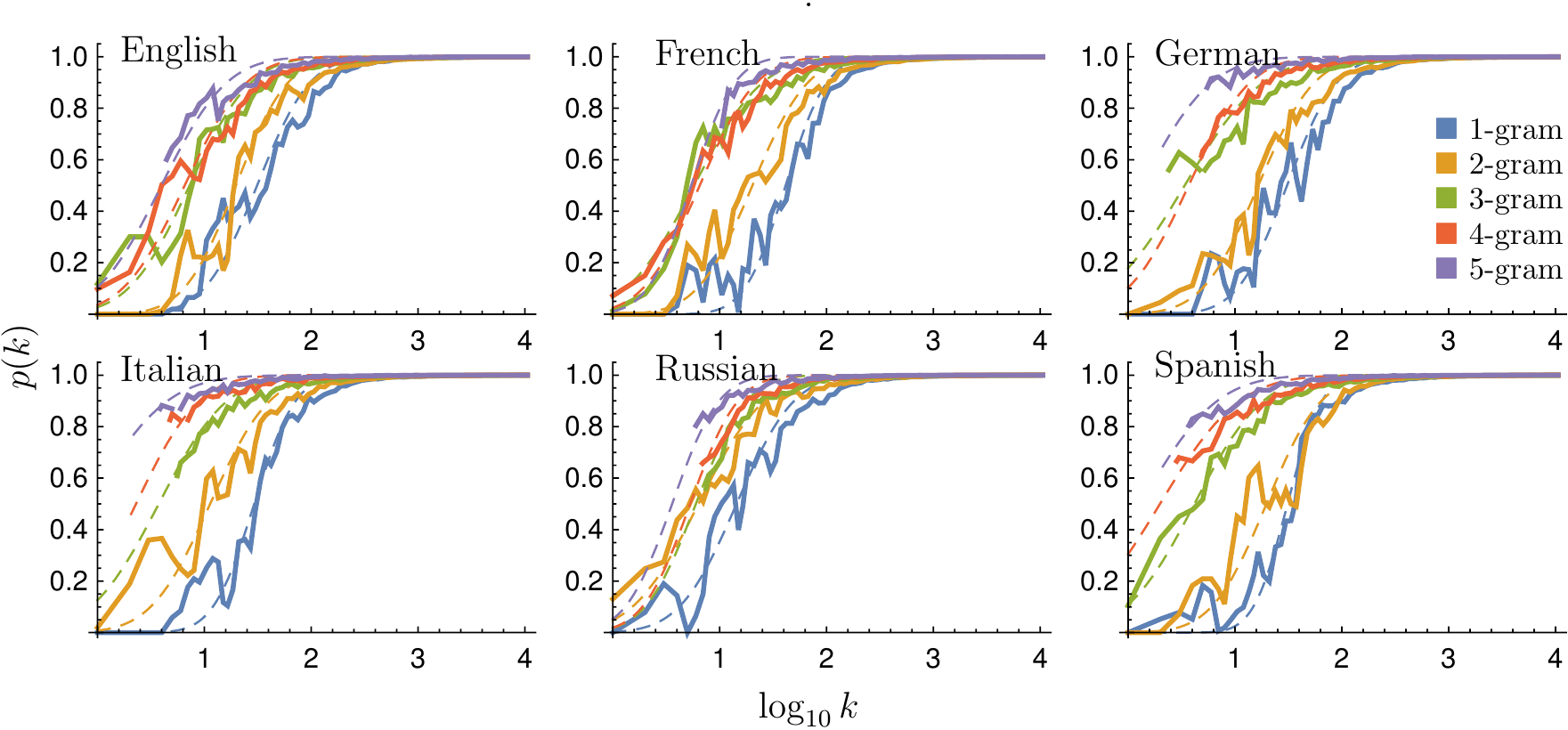}
\end{center}
\caption{
{\bf Change probability for different languages and $N$-grams}. Binned change probability $p(k)$ as a
function of rank $k$ for all languages and $N$ values considered (continuous lines). We also include fits according to the sigmoid in \eref{eq:sigmoid} (dashed lines), as in \Fref{fig:d}.
}
\label{fig:p}
\end{figure}

\begin{figure} 
\begin{center}
\includegraphics{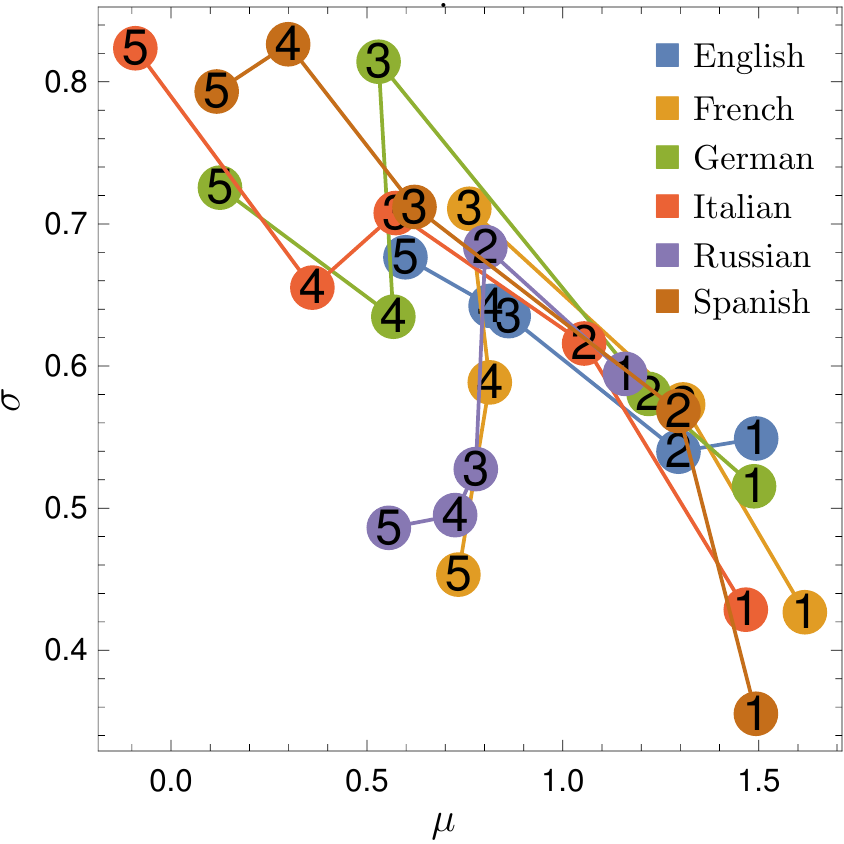}
\end{center}
\caption{{\bf Fitted parameters for change probability}. Parameters $\mu$ and $\sigma$ for the sigmoid fit of the change probability $p(k)$, for all languages (indicated by colors) and $N$ values (indicated by numbers). }
\label{fig:clustersP}
\end{figure}

Figure~\ref{fig:p} shows the behavior of the change probability $p(k)$ for all languages studied. We see that $p(k)$ grows faster than $d(k)$ for increasing rank $k$. The curves can also be well fitted with the sigmoid of \eref{eq:sigmoid} (fit parameters in \Tref{tab:changeprob_params}). Figure~\ref{fig:clustersP} shows the relationship between $\mu$ and $\sigma$ of the sigmoid fits for the change probability $p(k)$. As with the rank diversity, $\mu$ decreases with $N$ for each language, except for French and German between 3-grams and 4-grams. However, the $\sigma$ values seem to have a low correlation with $N$. We also analyze the difference between rank diversity and change probability, $d(k)-p(k)$ (Figure S1). As the change probability grows faster with rank $k$, the difference becomes negative and then grows together with the rank diversity. For large $k$, both rank diversity and change probability tend to one, so their difference is zero.

\subsection{Rank Entropy of $N$-gram usage}

\begin{figure} 
\begin{center}
\includegraphics{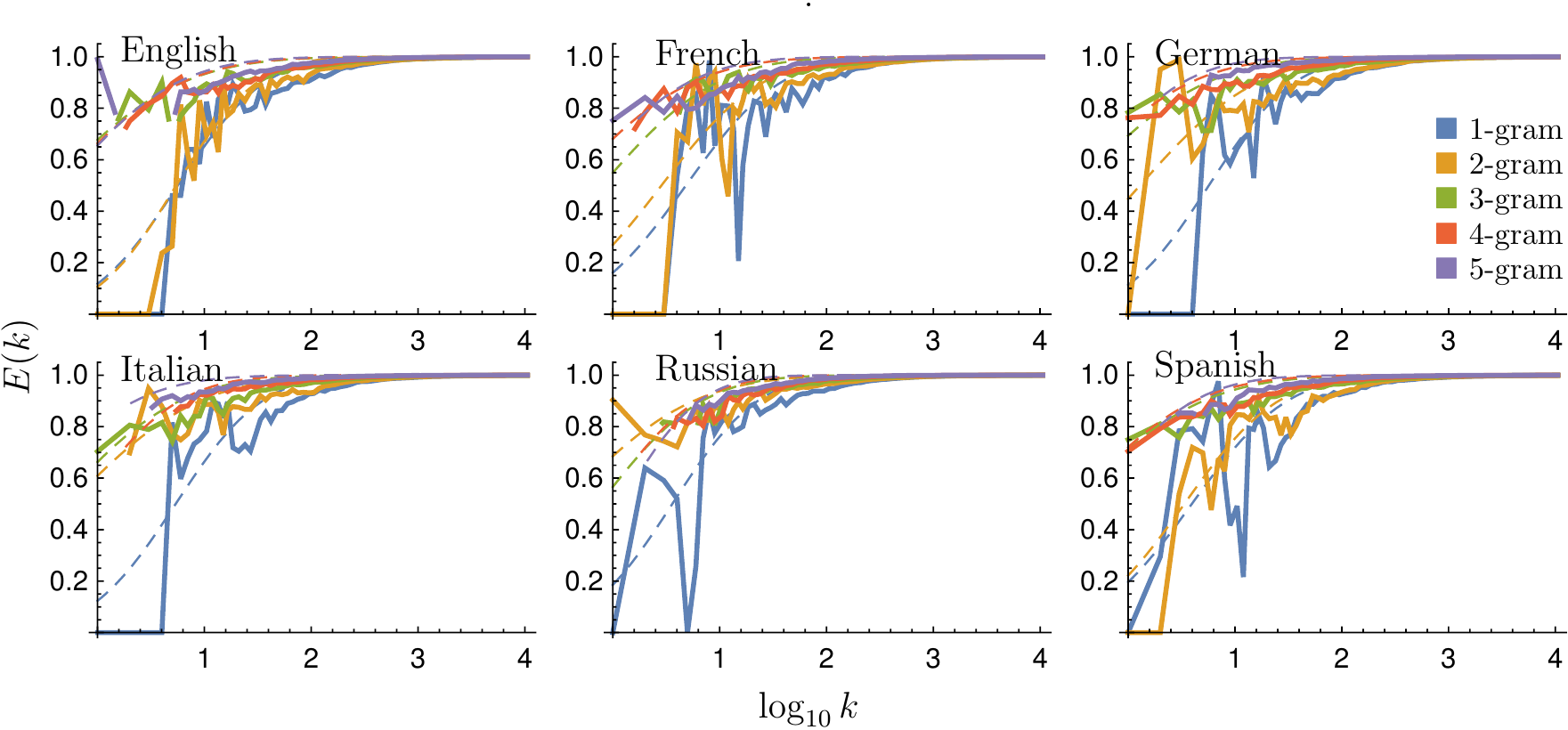}
\end{center}
\caption{{\bf Rank entropy for different languages and $N$-grams}. Binned rank entropy $E(k)$ as a
function of rank $k$ for all languages and $N$ values considered (continuous lines). We also include fits according to the sigmoid in \eref{eq:sigmoid} (dashed lines), as in \Fref{fig:d}.}
\label{fig:e}
\end{figure}

We can define another related measure: the rank entropy $E(k)$. Based on Shannon's information, it is simply the normalized information for the elements appearing at rank $k$ during all time intervals (see Methods). For example, if at rank $k=1$ only two $N$-grams appear, $d(1)=2/155$. Information is maximal when the probabilities of elements are homogeneous, \emph{i.e.} when each $N$-gram appears half of the time, as it is uncertain which of the elements will occur in the future. However, if one element appears only once, information will be minimal, as there will be a high probability that the other element will appear in the future.  As with the rank diversity and change probability, the rank entropy $E(k)$ also increases its value with rank $k$, even faster in fact, as shown in Figure~\ref{fig:e}. Similarly, $E(k)$ tends to be higher as $N$ grows, and may be fitted by the sigmoid of \eref{eq:sigmoid} at least for high enough $k$ (see fit parameters in \Tref{tab:rankent_params}) Notice that since rank entropy in some cases has already high values at $k=1$, the sigmoids can have negative $\mu$ values.

\begin{figure} 
\begin{center}
\includegraphics{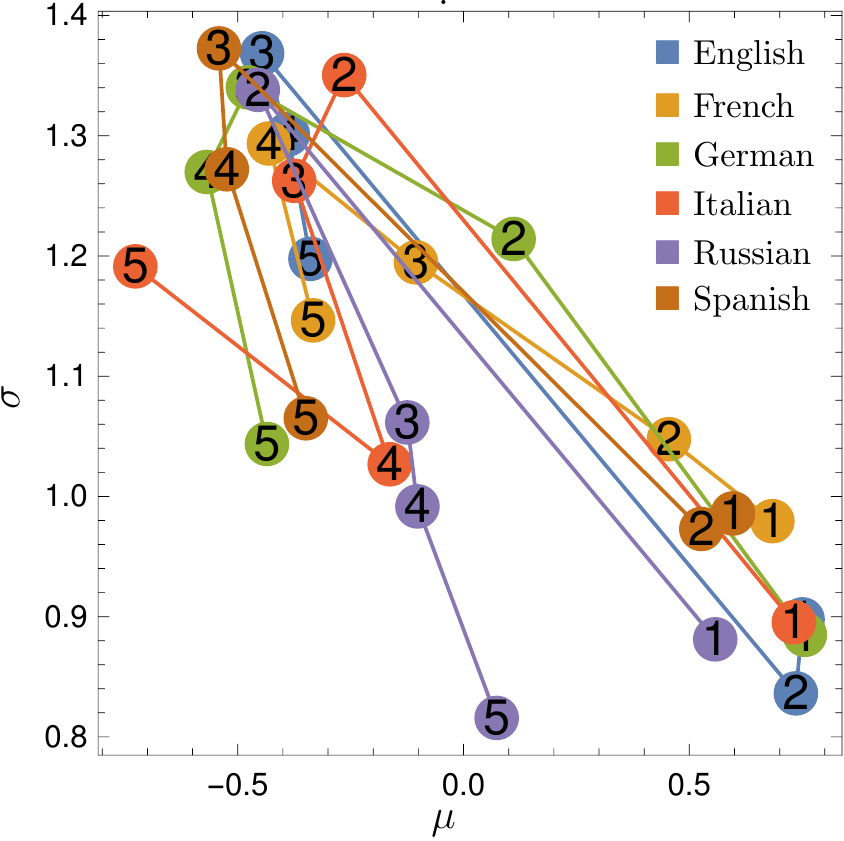}
\end{center}
\caption{{\bf Fitted parameters for rank entropy}. Parameters $\mu$ and $\sigma$ for the sigmoid fit of rank entropy $E(k)$, for all languages (indicated by colors) and $N$ values (indicated by numbers).}
\label{fig:clustersE}
\end{figure}

The $\mu$ and $\sigma$ values are compared in Figure~\ref{fig:clustersE}. The behavior of these parameters is more diverse than for rank diversity and change probability. Still, the curves tend to have a ``horseshoe'' shape, where $\mu$ decreases and $\sigma$ increases up to $N \approx 3$, and then $\mu$ slightly increases while $\sigma$ decreases.

\subsection{Rank Complexity of $N$-gram usage}

\begin{figure} 
\begin{center}
\includegraphics{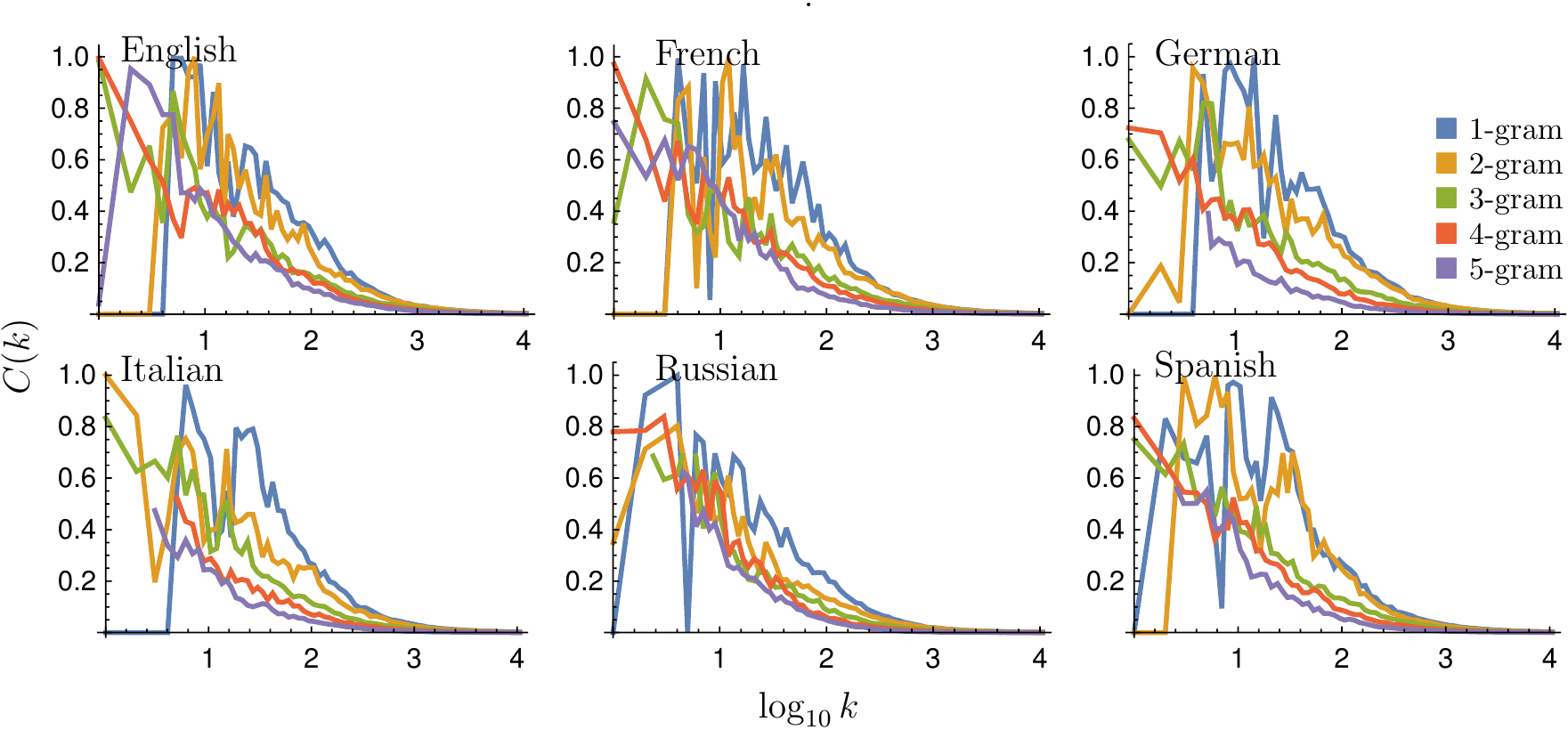}
\end{center}
\caption{{\bf Rank complexity for different languages and $N$-grams}. Binned rank complexity $C(k)$ as a
function of rank $k$ for all languages and $N$ values considered. Rank complexity tends to be greater for lower $N$, as rank entropy increases with $N$.}
\label{fig:c}
\end{figure}

Finally, we define the rank complexity $C(k)$ as
\begin{equation}
C(k)=4 E(k) (1-E(k)).
\label{eq:C}
\end{equation}
This measure of complexity represents a balance between stability (low entropy) and change (high entropy)~\citep{GershensonFernandez:2012,Fernandez2013Information-Mea,CxContinuous2016}. So complexity is minimal for extreme values of the normalized entropy [$E(k)=0$ or $E(k)=1$] and maximal for intermediate values [$E(k)=0.5$]. Figure~\ref{fig:c} shows the behaviour of the rank complexity $C(k)$ for all languages studied. In general, since $E(k)\approx0.5$ for low ranks, the highest $C(k)$ values appear for low ranks and decrease as $E(k)$ increases. $C(k)$ also decreases with $N$. Moreover, $C(k)$ curves reach values close to zero when $E(k)$ is close to one: around $k = 10^2$ for $N=5$ and $k = 10^3$ for $N=1$, for all languages.

\section{Discussion} 

Our statistical analysis suggests that human language is an example of a cultural construct
where macroscopic statistics (usage frequencies of $N$-grams for $N > 1$) cannot be deduced from microscopic
statistics (1-grams). Since not all word combinations are valid in the grammatical sense, in order to
study higher-order $N$-grams, the statistics of 1-grams are not enough, as shown by the null model results. In other words, $N$-gram statistics cannot be reduced to word statistics. This implies that multiple scales should be studied at the same time to understand language structure and use in a more integral fashion. We conclude not only that semantics and grammar cannot be reduced to syntax, but that even within syntax, higher scales ($N$-grams with $N > 1$) have an emergent, relevant structure which cannot be exclusively deduced from the lowest scale ($N = 1$).

While the alphabet, the grammar, and the subject matter of a text can vary greatly among languages, unifying statistical patterns do exist, and they allow us to study language as a social and cultural phenomenon without limiting our conclusions to one specific language. We have shown that despite many clear differences between the six languages we have studied, each language balances a versatile but stable core of words with less frequent but adaptable (and more content-specific) words in a very similar way. This leads to linguistic structures that deviate far from what would be expected in a random `language' of shuffled 1-grams. In particular, it causes the most commonly used word combinations to deviate further from random that those at the other end of the usage scale.

If we are to assume that all languages have converged on the same pattern because it is in some way `optimal', then it is perhaps this  statistical property that allows word combinations to carry more information that the sum of their parts; to allow words to combine in the most efficient way possible in order to convey a concept that cannot be conveyed through a sequence of disconnected words. The question of whether or not the results we report here are consistent with theories of language evolution \citep{Nowak8028,Cancho788,1742-5468-2006-06-P06014} is certainly a topic for discussion and future research.


Apart from studying rank diversity, in this work we have introduced measures of change probability, rank entropy, and rank complexity. Analytically, the change probability is simpler to treat than rank diversity, as the latter varies with the number of time intervals considered ($T$), while the former is more stable (for a large enough number of observations). Still, rank diversity produces smoother curves and gives more information about rank dynamics, since the change probability grows faster with $k$. Rank entropy grows even faster, but all three measures [$d(k)$, $p(k)$, and $E(k)$] seem related, as they tend to grow with $k$ and $N$ in a similar fashion. Moreover, all three measures can be relatively well fitted by sigmoid curves (the worst fit has $e=0.02$, as seen in \Tref{tab:rankdiv_params}-\ref{tab:rankent_params}). Our results suggests that a sigmoid functional shape fits rank diversity the best for low ranks, as the change probability and rank entropy have greater variability in that region.

In \citet{Cocho2015}, we used the parameters of the sigmoid fit to rank diversity as an approximation of language core size, \emph{i.e.} the number of 1-grams minimally required to speak a language. Assuming that these basic words are frequently used (low $k$) and thus have $d(k)<1$, we consider the core size to be bounded by $\log_{\mathrm{10}} k = \mu+2\sigma$. As \Tref{tab:core_params} shows, this value decreases with $N$, \emph{i.e.} $N$-gram structures with larger $N$ tend to have smaller cores. However, if the number of different words found on cores are counted, they increase from monograms to digrams, except for Spanish and Italian. From $N=2$, the number of words in cores decreases constantly for all languages. This suggests that core words can be combined to form more complex expressions without the requirement of learning new words. English and French tend to have more words in their cores, while Russian has the least. 
It is interesting to note that the null model produces cores with about twice as many words as real 2-grams. Also, only in language cores rank complexity values are not close to zero. In other words, only ranks within the core have a high rank complexity.

Our results may have implications for next-word prediction algorithms used in modern typing interfaces like smartphones. Lower ranked $N$-grams tend to be more predictable (higher $z$-scores and lower next word entropy on average). Thus, next-word prediction should adjust the $N$ value (scale) depending on the expected rank of the recent, already-typed words. If these are not in top ranked $N$-grams, then $N$ should be decreased. For example, on the iOS 11 platform, after typing {\it `United States of'}, the system suggests {\it `the'}, {\it `all'}, and {\it `a'}, as the next-word prediction by analyzing 2-grams. However, it is clear that the most probable next-word is {\it `America'}, as this is a low-ranked 4-gram. 

Beyond the previous considerations, perhaps the most relevant aspect of our results is that the rank dynamics of language use is generic not only for all six languages, but for all five scales studied. Whether the generic properties of rank diversity and related measures are universal still remains to be explored. Yet, we expect this and other research questions to be answered in the coming years as more data on language use and human culture becomes available.

\section{Methods} 

\subsection{Data description}
\label{data}

Data was obtained from the Google Books $N$-gram dataset~\footnote{\url{http://storage.googleapis.com/books/ngrams/books/datasetsv2.html}}, filtered and processed to obtain ranked $N$-grams for each year for each language. Data considers only the first $11,140$ ranks, as this was the maximum rank available for all time intervals and languages studied.
From these, rank diversity, change probability, rank entropy, and rank complexity were calculated as follows. Rank diversity is given by
\begin{equation}
d(k)=\frac{|X(k)|}{T},
\end{equation}
where $|X(k)|$ is the cardinality (\emph{i.e.} number of elements) that appear at rank $k$ during all $T=155$ time intervals (between 1855 and 2009 with one-year differences, or $\Delta t = 1$). The change probability is
\begin{equation}
p(k)=\frac{\sum_{t=0}^{t=T-1}{1-\delta(X(k,t), X(k,t+1))}}{T -1},
\end{equation}
where $\delta(X(k,t), X(k,t+1))$ is the Kronecker delta; equal to zero if there is a change of $N$-gram in rank $k$ in $\Delta t$ [\emph{i.e.} the element $X(k,t)$ is different from element $X(k,t+1)$], and equal to one if there is no change. The rank entropy is given by
\begin{equation}
E(k)=-\kappa \sum_{i=1}^{|X(k)|} p_i \log p_i,
\label{eq:E}
\end{equation}
where 
\begin{equation}
\kappa = \frac{1}{\log_2 |X(k)|},
\end{equation}
so as to normalize $E(k)$ in the interval $[0,1]$. Note that $|X(k)|$ is the alphabet length, \emph{i.e.} the number of elements that have occurred at rank $k$. Finally, the rank complexity is calculated using Eq.~\ref{eq:C} and Eq.~\ref{eq:E}~\citep{Fernandez2013Information-Mea}. 

\subsection{Modelling shuffled data}
\label{null}

We first describe a shuffling process that eliminates any structure found within the 2-gram data, while preserving the frequency of individual words. Consider a sequence consisting of the most frequent word a number of times equal to its frequency, followed by the second most frequent word a number of times equal to its frequency, and so on all the way up to the $11,140^\mathrm{th}$ most frequent word (i.e until all the words in the monogram data have been exhausted). Now suppose we shuffle this sequence and obtain the frequencies of 2-grams in the new sequence. Thus, we have neglected any grammatical rules about which words are allowed to follow which others (we can have the same word twice in the same 2-gram, for example), but the frequency of words remains the same.

We now derive an expression for the probability that a 2-gram will have a given frequency after shuffling has been performed. Let $f_{i}$ denote the number of times the word $i$ appears in the text, and $f_{ij}$ the number of times the 2-gram $ij$ appears. Additionally, $F=\sum_{i}f_{i}$. We want to know the probability $P(f_{ij})$ that $ij$ appears exactly $f_{ij}$ times in the table. We can think of $P(f_{ij})$ as the probability that exactly $f_{ij}$ occurrences of $i$ are followed by $j$. Supposing $f_{i}<f_{j}$, $f_{ij}$ is determined by $f_{i}$ independent Bernoulli trials with the probability of success equal to the probability that the next word will be $j$, \emph{i.e.} $f_{j}/F$. In this case we have 
\begin{equation}
P(f_{ij})=\binom{f_{i}}{f_{ij}}\left(\frac{f_{j}}{F}\right)^{f_{ij}}\left(1-\frac{f_{j}}{F}\right)^{f_{i}-f_{ij}}.
\end{equation}
This distribution meets the condition that allows it to be approximated by a Poisson distribution, namely that $f_{i}f_{j}/F$ is constant, so we have 
\begin{equation}
P(f_{ij})\approx\frac{\lambda_{ij}^{f_{ij}}e^{-\lambda_{ij} }}{f_{ij}!},
\end{equation}
where 
\begin{equation}
\lambda_{ij}=\frac{f_{i}f_{j}}{F}
\end{equation}
is the mean, and also the variance, of the distribution of values of $f_{ij}$.

For each 2-gram we calculate the $z$-score. This is a normalized frequency of its occurrence, \emph{i.e.} we normalize the actual frequency $f_{ij}$ by subtracting the mean of the null distribution and dividing by the standard deviation,
\begin{equation}
z_{ij}=\frac{f_{ij}-\mu_{ij}}{\sigma_{ij}}=\frac{f_{ij}-\lambda_{ij}}{\sqrt{\lambda_{ij} }}.
\end{equation}
In other words, the $z$-score tells us how many standard deviations the actual frequency is from the mean of the distribution derived from the shuffling process. The result is that the 2-grams with the highest $z$-scores are those which occur relatively frequently but their component words occur relatively infrequently.

\textbf{Normalization.} To compare $z$-scores of different languages, we normalize to eliminate the effects of incomplete data. Specifically, we normalize $z$-scores by dividing by the upper bound (which happens to be equal in order of magnitude to the lower bound). The highest possible $z$-score occurs in cases where $f_{i}=f_{j}=f_{ij}=f$. Therefore $\lambda_{ij}=f^{2}/F$ and
\begin{equation}
z_{ij}\leq\sqrt{F}\left(1-\frac{f}{F}\right)<\sqrt{F},
\end{equation}
so an upper bound exists at $\sqrt{F}$. Similarly, The lowest possible $z$-score would hypothetically occur when $f_{i}=f_{j}\approx F/2$ and $f_{ij}=f$, giving
\begin{equation}
z_{ij}\leq\frac{\sqrt{F}}{F}\left(2f-\frac{F}{2}\right)>-\frac{\sqrt{F}}{2}.
\end{equation}
We thus define the normalized $z$-score as
\begin{equation}
\label{normalized_z}
\hat{z}_{ij}=\frac{z_{ij}}{\sqrt{F}}.
\end{equation}

\textbf{The relationship between rank and $z$-score.} To understand how the $z$-score changes as a function of rank, we look at another special case: suppose that $i$ is a word that is found to be the first word in a relatively large number of 2-grams, and that all occurrences of the word $j$ are preceded by $i$. In such cases we have $f_{i,j}=f_{j}$, so Eq.\eqref{normalized_z} reduces to
\begin{equation}
\hat{z}_{ij}=\left(\frac{1}{f_{i}}-\frac{1}{F}\right)(f_{i}f_{j})^{1/2}.
\end{equation}
Now consider only the subset of 2-grams that start with $i$ and end with words that are only ever found to be preceded by $i$. Since $f_{i}$ is constant within this subset, we have $\hat{z}_{ij}=Af_{j}^{1/2}$, where $A$ is a constant. If we now assume that Zipf's law holds for the set of second words in the subset, \emph{i.e.} that $f_{j}=Br_{j}^{-1}$ where $r_{j}$ is the rank of $j$ and $B$ another constant, then we have $\hat{z}_{ij}=Cr_{j}^{-1/2}$, with $C$ a third constant.

\textbf{Data.} Unlike in other parts of this study, the shuffling analysis is applied to the $10^{5}$ lowest ranked 2-grams.

\subsection{Next-word entropy}

The relationship between rank and $z$-score of 2-grams appears to be, at least partially, a consequence of the existence of high frequency core words that can be followed by many possible next words. This diversity of next words can be quantified by what we call the next-word entropy. Given a word $i$, we define the next-word entropy, $E_{i}^{\text{nw}}$, of $i$ to be the (non-normalized) Shannon entropy of the distribution of 2-gram frequencies of 2-grams that have $i$ as the first word,
\begin{equation}
E_{i}^{\text{nw}}=-\sum_{i}\frac{f_{ij}}{f_{i}}\log\left(\frac{f_{ij}}{f_{i}}\right).
\label{eq:Enw}
\end{equation}

\subsection{Fitting process}

The curve fitting for rank diversity, change probability, and rank entropy has been made with the scipy-numpy package using the non-linear least squares method (Levenberg-Marquardt algorithm). For rank entropy, we average data over each ten ranks, $\overline{k}_i =
\frac{\sum_{i=0}^{n/10} k_{i}}{10}$, as well as over rank entropy values,
$\overline{E(k_i)} = \frac{\sum_{i=0}^{n/10} E(k_{i})}{10}$. With this averaged data, we adjust a cumulative normal (\emph{erf} function) over the data of $\log_{10}(\overline{k}_i)$ and $\overline{E(k_i)}$. For rank diversity and change probability, we average data over points equally spaced in $\log_{10}(k_i)$. Like for rank entropy, a sigmoid (Eq.~\ref{eq:sigmoid}) is fitted for $\log_{10}(k)$ and $d(k)$, as well as for $\log_{10}(k)$ and $p(k)$. To calculate the mean quadratic error, we use 
\begin{equation}
e = \sqrt{\frac{\sum_{i=1}^{n}(\hat{X_i}-X_i)^2}{n}}, 
\end{equation}
where $\hat{X_i}$ is the value of the sigmoid adjusted to rank $k_i$ and $X_i$ is the real value of $d(k_i)$. For $p(k)$ and $E(k)$ the error is calculated in the same way.

\section*{Conflict of Interest Statement} 

The authors declare that the research was conducted in the absence of any commercial or financial relationships that could be construed as a potential conflict of interest.
\section*{Author Contributions} 

All authors contributed to the conception of the paper. JAM, EC, and SS processed and analysed the data. EC and GI devised the null model. CP and EC made the figures. EC, GI, JF, and CG wrote sections of the paper. All authors contributed to manuscript revision, read and approved the final version of the article.


\section*{Funding} 
We acknowledge support from UNAM-PAPIIT Grant No. IG100518, CONACyT
Grant No. 285754, and the Fundaci\'on Marcos Moshinsky.


\section*{Supplemental Data} 

Additional Tables and Figures.

\bibliographystyle{frontiersinSCNS_ENG_HUMS} 
\bibliography{ngrams}

\clearpage

\section*{Figure captions} 

\clearpage

\section*{Tables} 

\begin{table}[h]
\centering
\resizebox{\columnwidth}{!}{ 
\begin{tabular}{|l|l|l|l|l|l|l|l|l|l|l|l|l|l|l|l|l|l|l|l|l|l|l|l|l|l|l|l|l|l|l|l|}
\hline
\multirow{2}{*}{} & \multicolumn{3}{c|}{\textbf{1grams}} & \multicolumn{3}{c|}{\textbf{2grams}} & \multicolumn{3}{c|}{\textbf{3grams}} & \multicolumn{3}{c|}{\textbf{4grams}} & \multicolumn{3}{c|}{\textbf{5grams}} & \multicolumn{3}{c|}{\textbf{Random 2grams}} \\ 
\cline{2-19} 
& $\mu$ & $\sigma$ & $e$ & $\mu$ & $\sigma$ & $e$ & $\mu$ & $\sigma$ & $e$ & $\mu$ & $\sigma$ & $e$ & $\mu$ & $\sigma$ & $e$ & $\mu$ & $\sigma$ & $e$ \\ 
\hline
\textbf{English} & 2.281 & 0.62 & 0.019 & 2.135 & 0.721 & 0.016 & 1.822 & 0.823 & 0.013 & 1.755 & 0.775 & 0.012 & 1.553 & 0.822 & 0.01 & 2.605 & 0.598 & 0.024 \\ 
\textbf{French} & 2.273 & 0.63 & 0.02 & 2.183 & 0.689 & 0.016 & 1.815 & 0.818 & 0.013 & 1.646 & 0.82 & 0.011 & 1.385 & 0.849 & 0.01 & 2.684 & 0.598 & 0.022 \\ 
\textbf{German} & 2.234 & 0.611 & 0.018 & 2.123 & 0.695 & 0.015 & 1.684 & 0.839 & 0.012 & 1.461 & 0.818 & 0.009 & 0.958 & 0.936 & 0.007 & 2.509 & 0.636 & 0.02 \\ 
\textbf{Italian} & 2.208 & 0.638 & 0.017 & 2.031 & 0.717 & 0.014 & 1.663 & 0.817 & 0.012 & 1.261 & 0.935 & 0.009 & 0.953 & 0.95 & 0.007 & 2.53 & 0.627 & 0.019 \\ 
\textbf{Russian} & 2.074 & 0.613 & 0.014 & 1.811 & 0.767 & 0.012 & 1.545 & 0.786 & 0.01 & 1.427 & 0.713 & 0.008 & 1.264 & 0.705 & 0.006 & 2.228 & 0.628 & 0.017 \\ 
\textbf{Spanish} & 2.137 & 0.69 & 0.017 & 2.076 & 0.675 & 0.017 & 1.701 & 0.843 & 0.012 & 1.379 & 0.904 & 0.01 & 1.025 & 0.96 & 0.008 & 2.573 & 0.551 & 0.024 \\ 
\hline 
\end{tabular}
}
\caption{{\bf Fit parameters for rank diversity for different languages, $N$-grams and null model}. Mean $\mu$, standard deviation $\sigma$, and error $e$ for the sigmoid fit of the rank diversity $d(k)$ according to \eref{eq:sigmoid}. We also show the fit parameters for the null model of \Fref{fig:shuffled}.}
\label{tab:rankdiv_params}
\end{table}

\begin{table}[h]
\centering
\resizebox{\columnwidth}{!}{ 
\begin{tabular}{|l|l|l|l|l|l|l|l|l|l|l|l|l|l|l|l|l|l|l|l|l|l|l|l|l|l|l|l|l|l|l|l|}
\hline
\multirow{2}{*}{} & \multicolumn{3}{c|}{\textbf{1grams}} & \multicolumn{3}{c|}{\textbf{2grams}} & \multicolumn{3}{c|}{\textbf{3grams}} & \multicolumn{3}{c|}{\textbf{4grams}} & \multicolumn{3}{c|}{\textbf{5grams}} \\ 
\cline{2-16} 
& $\mu$ & $\sigma$ & $e$ & $\mu$ & $\sigma$ & $e$ & $\mu$ & $\sigma$ & $e$ & $\mu$ & $\sigma$ & $e$ & $\mu$ & $\sigma$ & $e$ \\ 
\hline
\textbf{English} & 1.494 & 0.549 & 0.008 & 1.295 & 0.54 & 0.009 & 0.862 & 0.635 & 0.006 & 0.816 & 0.642 & 0.005 & 0.598 & 0.677 & 0.005 \\ 
\textbf{French} & 1.618 & 0.427 & 0.009 & 1.307 & 0.573 & 0.008 & 0.761 & 0.711 & 0.006 & 0.814 & 0.589 & 0.004 & 0.733 & 0.453 & 0.004 \\ 
\textbf{German} & 1.488 & 0.515 & 0.009 & 1.219 & 0.58 & 0.007 & 0.53 & 0.814 & 0.004 & 0.567 & 0.635 & 0.003 & 0.125 & 0.726 & 0.003 \\ 
\textbf{Italian} & 1.467 & 0.429 & 0.009 & 1.054 & 0.616 & 0.007 & 0.574 & 0.708 & 0.004 & 0.361 & 0.655 & 0.004 & -0.091 & 0.824 & 0.002 \\ 
\textbf{Russian} & 1.159 & 0.595 & 0.007 & 0.802 & 0.684 & 0.005 & 0.778 & 0.527 & 0.004 & 0.725 & 0.495 & 0.003 & 0.556 & 0.486 & 0.003 \\ 
\textbf{Spanish} & 1.493 & 0.355 & 0.009 & 1.295 & 0.568 & 0.009 & 0.621 & 0.712 & 0.004 & 0.299 & 0.826 & 0.003 & 0.116 & 0.793 & 0.003 \\ 
\hline 
\end{tabular}
}
\caption{{\bf Fit parameters for change probability for different languages}. Mean $\mu$, standard deviation $\sigma$, and error $e$ for the sigmoid fit of the change probability $p(k)$ according to \eref{eq:sigmoid}. }
\label{tab:changeprob_params}
\end{table}

\begin{table}[h]
\centering
\resizebox{\columnwidth}{!}{ 
\begin{tabular}{|l|l|l|l|l|l|l|l|l|l|l|l|l|l|l|l|l|l|l|l|l|l|l|l|l|l|l|l|l|l|l|l|}
\hline
\multirow{2}{*}{} & \multicolumn{3}{c|}{\textbf{1grams}} & \multicolumn{3}{c|}{\textbf{2grams}} & \multicolumn{3}{c|}{\textbf{3grams}} & \multicolumn{3}{c|}{\textbf{4grams}} & \multicolumn{3}{c|}{\textbf{5grams}} \\ 
\cline{2-16} 
& $\mu$ & $\sigma$ & $e$ & $\mu$ & $\sigma$ & $e$ & $\mu$ & $\sigma$ & $e$ & $\mu$ & $\sigma$ & $e$ & $\mu$ & $\sigma$ & $e$ \\ 
\hline
\textbf{English} & 0.751 & 0.898 & 0.009 & 0.736 & 0.836 & 0.009 & -0.446 & 1.369 & 0.003 & -0.389 & 1.301 & 0.002 & -0.339 & 1.198 & 0.004 \\ 
\textbf{French} & 0.684 & 0.979 & 0.012 & 0.455 & 1.048 & 0.01 & -0.105 & 1.195 & 0.006 & -0.431 & 1.294 & 0.002 & -0.333 & 1.146 & 0.002 \\ 
\textbf{German} & 0.756 & 0.885 & 0.009 & 0.111 & 1.214 & 0.007 & -0.478 & 1.34 & 0.003 & -0.569 & 1.27 & 0.001 & -0.435 & 1.044 & 0.001 \\ 
\textbf{Italian} & 0.731 & 0.895 & 0.01 & -0.264 & 1.35 & 0.004 & -0.375 & 1.263 & 0.002 & -0.163 & 1.027 & 0.003 & -0.727 & 1.191 & 0.001 \\ 
\textbf{Russian} & 0.557 & 0.881 & 0.009 & -0.456 & 1.338 & 0.003 & -0.125 & 1.062 & 0.003 & -0.102 & 0.992 & 0.002 & 0.073 & 0.816 & 0.002 \\ 
\textbf{Spanish} & 0.597 & 0.986 & 0.011 & 0.527 & 0.973 & 0.008 & -0.542 & 1.373 & 0.002 & -0.524 & 1.272 & 0.002 & -0.349 & 1.065 & 0.001 \\ 
\hline 
\end{tabular}
}
\caption{{\bf Fit parameters for rank entropy for different languages}. Mean $\mu$, standard deviation $\sigma$, and error $e$ for the sigmoid fit of the rank entropy $E(k)$ according to \eref{eq:sigmoid}. }
\label{tab:rankent_params}
\end{table}

\begin{table}[h]
\centering
\resizebox{\columnwidth}{!}{ 
\begin{tabular}{|l|l|l|l|l|l|l|l|l|l|l|l|l|l|l|l|l|l|l|l|l|l|l|l|l|l|l|l|l|l|l|l|}
\hline
\multirow{2}{*}{} & \multicolumn{2}{c|}{\textbf{1grams}} & \multicolumn{2}{c|}{\textbf{2grams}} & \multicolumn{2}{c|}{\textbf{3grams}} & \multicolumn{2}{c|}{\textbf{4grams}} & \multicolumn{2}{c|}{\textbf{5grams}} & \multicolumn{2}{c|}{\textbf{Random 2grams}} \\ 
\cline{2-13} 
& $\mu + 2 \sigma$ & \# words & $\mu + 2 \sigma$ & \# words & $\mu + 2 \sigma$ & \# words &  $\mu + 2 \sigma$ & \# words & $\mu + 2 \sigma$ & \# words & $\mu + 2 \sigma$ & \# words\\ 
\hline
\textbf{English} & 3.521 & 3004 & 3.578 & 3553 & 3.468 & 2813 & 3.304 & 1949 & 3.196 & 1512 & 3.801 & 6322 \\ 
\textbf{French} & 3.534 & 3173 & 3.561 & 3469 & 3.45 & 2705 & 3.287 & 1830 & 3.083 & 1134 & 3.881 & 7601 \\ 
\textbf{German} & 3.456 & 2622 & 3.512 & 3057 & 3.362 & 2198 & 3.097 & 1203 & 2.83 & 662 & 3.78 & 6032 \\ 
\textbf{Italian} & 3.485 & 2856 & 3.465 & 2815 & 3.297 & 1919 & 3.131 & 1303 & 2.854 & 696 & 3.784 & 6078 \\ 
\textbf{Russian} & 3.3 & 1827 & 3.345 & 1997 & 3.118 & 1216 & 2.853 & 681 & 2.674 & 452 & 3.483 & 3042 \\ 
\textbf{Spanish} & 3.516 & 2988 & 3.426 & 2529 & 3.386 & 2339 & 3.187 & 1479 & 2.945 & 842 & 3.675 & 4728 \\ 
\hline 
\end{tabular}
}
\caption{{\bf Language core parameters.} Upper bound rank $\log_{\mathrm{10}} k = \mu + 2 \sigma$ for the estimated core size of all languages studied, according to the sigmoid fit of \eref{eq:sigmoid}, as well as the number of words included in the $N$-grams within the core in the year 2009.} 
\label{tab:core_params}
\end{table}




\section*{Supplementary material}

frontiers

## *Supplementary Material*:
## Rank Dynamics of Word Usage at Multiple Scales

# 1 SUPPLEMENTARY TABLES AND FIGURES

## 1.1 Figures

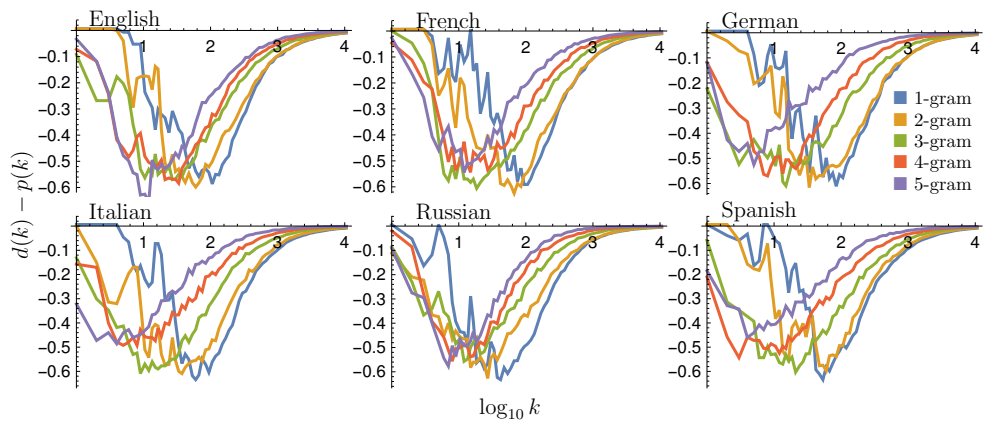

Figure S1: **Difference between rank diversity and change probability across languages and $N$-grams**. As $p(k)$ grows faster than $d(k)$, these curves decrease their value to a minimum close to 0.6, for then increasing when $d(k)$ starts growing and become zero when both measures reach their maximum value of one. Windowing is done averaging $d(k)$ every 0.05 in $log k$.

## 1.2 Tables

| Rank | 1700 | 1800 | 1900 | 2000 |
|------|------|------|------|------|
| 1 | the | the | the | the |
| 2 | of | of | of | of |
| 3 | and | and | and | and |
| 4 | to | to | to | to |
| 5 | in | in | in | a |
| 6 | a | a | a | in |
| 7 | is | that | that | that |
| 8 | his | is | is | is |
| 9 | he | i | was | for |
| 10 | that | his | as | the |
| 11 | for | it | i | was |
| 12 | not | was | his | with |
| 13 | as | with | with | as |
| 14 | i | he | it | i |
| 15 | this | as | for | not |
| 16 | it | for | the | on |
| 17 | ' | be | he | 's |
| 18 | by | not | by | it |
| 19 | be | by | be | be |
| 20 | and | which | not | by |
| 21 | their | the | which | or |
| 22 | was | on | on | are |
| 23 | surveyor | at | at | you |
| 24 | with | this | had | from |
| 25 | upon | had | or | at |
| 26 | but | from | from | his |
| 27 | no | have | are | he |
| 28 | are | their | have | have |
| 29 | have | but | this | an |
| 30 | were | you | ' | this |

**Table S1.** Top English 1-grams for arbitrary years.

| Rank | **1700** | **1800** | **1900** | **2000** |
|---|---|---|---|---|
| 1  | of–the       | of–the     | of–the    | of–the    |
| 2  | in–the       | in–the     | in–the    | in–the    |
| 3  | to–the       | to–the     | to–the    | to–the    |
| 4  | the–surveyor | and–the    | and–the   | on–the    |
| 5  | and–the      | to–be      | on–the    | and–the   |
| 6  | upon–the     | on–the     | to–be     | for–the   |
| 7  | for–the      | by–the     | by–the    | to–be     |
| 8  | that–the     | of–his     | of–a      | of–a      |
| 9  | by–the       | from–the   | for–the   | from–the  |
| 10 | of–his       | with–the   | from–the  | with–the  |
| 11 | to–be        | of–a       | with–the  | at–the    |
| 12 | the–stage    | for–the    | at–the    | that–the  |
| 13 | in–his       | it–is      | that–the  | in–a      |
| 14 | to–make      | at–the     | it–is     | by–the    |
| 15 | with–the     | that–the   | of–his    | as–a      |
| 16 | is–not       | in–a       | in–a      | is–a      |
| 17 | the–surveyor | all–the    | with–a    | is–the    |
| 18 | the–drama    | is–the     | the–same  | it–is     |
| 19 | as–the       | with–a     | is–the    | do–not    |
| 20 | of–their     | it–was     | it–was    | with–a    |
| 21 | not–to       | in–his     | is–a      | did–not   |
| 22 | does–not     | i–have     | it–is     | can–be    |
| 23 | at–the       | as–the     | as–the    | as–the    |
| 24 | and–that     | he–was     | have–been | to–a      |
| 25 | in–their     | have–been  | as–a      | the–same  |
| 26 | he–is        | that–he    | he–was    | for–a     |
| 27 | from–the     | the–same   | had–been  | is–not    |
| 28 | the–reader   | of–their   | may–be    | into–the  |
| 29 | that–the     | had–been   | all–the   | it–was    |
| 30 | out–of       | of–this    | that–he   | was–a     |

**Table S2.** Top English 2-grams for arbitrary years.

| Rank | **1700** | **1800** | **1900** | **2000** |
|------|----------|----------|----------|----------|
| 1 | so–much–as | one–of–the | one–of–the | one–of–the |
| 2 | of–the–stage | as–well–as | the–united–states | as–well–as |
| 3 | the–surveyor–'s | part–of–the | part–of–the | the–united–states |
| 4 | by–the–surveyor | i–do–not | i–do–not | i–do–not |
| 5 | as–the–surveyor | out–of–the | as–well–as | part–of–the |
| 6 | up–to–the | the–name–of | out–of–the | the–end–of |
| 7 | the–rest–of | in–order–to | can–not–be | out–of–the |
| 8 | the–english–stage | i–can–not | and–in–the | in–order–to |
| 9 | as–well–as | can–not–be | in–order–to | end–of–the |
| 10 | to–this–i | the–same–time | the–end–of | some–of–the |
| 11 | part–of–the | and–in–the | in–which–the | the–number–of |
| 12 | of–the–fable | of–the–world | of–the–united | to–be–a |
| 13 | not–to–be | according–to–the | that–he–was | i–did–not |
| 14 | not–so–much | that–it–was | is–to–be | be–able–to |
| 15 | of–the–play | of–the–most | i–can–not | the–use–of |
| 16 | of–the–drama | that–he–was | of–the–most | a–number–of |
| 17 | comedy–and–tragedy | is–to–be | the–same–time | can–not–be |
| 18 | but–the–surveyor | the–united–states | that–of–the | the–fact–that |
| 19 | upon–the–stage | at–the–same | that–it–was | in–the–united |
| 20 | tragedy–and–comedy | it–is–not | the–name–of | do–not–know |
| 21 | the–surveyor–is | to–have–been | some–of–the | the–same–time |
| 22 | the–reader–may | that–of–the | of–the–world | in–terms–of |
| 23 | the–authority–of | that–it–is | that–he–had | in–which–the |
| 24 | rest–of–the | that–he–had | the–fact–that | there–is–a |
| 25 | one–would–think | of–all–the | in–the–same | a–lot–of |
| 26 | is–not–so | to–be–a | side–of–the | there–is–no |
| 27 | and–as–for | not–to–be | it–is–not | i–can–not |
| 28 | a–man–'s | in–the–same | that–it–is | the–rest–of |
| 29 | to–no–purpose | in–the–world | to–be–a | you–do–not |
| 30 | to–make–the | there–is–no | at–the–same | side–of–the |

**Table S3.** Top English 3-grams for arbitrary years.

| Rank | 1700 | 1800 | 1900 | 2000 |
|---|---|---|---|---|
| 1 | not–so–much–as | at–the–same–time | of–the–united–states | in–the–united–states |
| 2 | to–this–i–answer | of–the–united–states | at–the–same–time | the–end–of–the |
| 3 | the–rest–of–the | at–the–head–of | the–end–of–the | i–do–not–know |
| 4 | this–may–serve–to | in–the–midst–of | one–of–the–most | at–the–end–of |
| 5 | the–authority–of–the | the–name–of–the | i–do–not–know | at–the–same–time |
| 6 | i–must–tell–him | in–the–county–of | on–the–part–of | of–the–united–states |
| 7 | but–it–seems–the | one–of–the–most | in–the–case–of | as–well–as–the |
| 8 | as–far–as–it | i–do–not–know | for–the–purpose–of | the–rest–of–the |
| 9 | and–this–may–serve | on–the–part–of | at–the–end–of | on–the–other–hand |
| 10 | a–great–part–of | as–well–as–the | is–one–of–the | one–of–the–most |
| 11 | a–great–deal–of | in–the–course–of | in–the–midst–of | as–a–result–of |
| 12 | whence–it–will–follow | the–rest–of–the | in–the–united–states | is–one–of–the |
| 13 | upon–the–english–stage | in–the–name–of | on–the–other–hand | in–the–case–of |
| 14 | upon–the–account–of | the–end–of–the | as–well–as–the | in–the–form–of |
| 15 | to–take–notice–of | the–head–of–the | for–the–first–time | do–not–want–to |
| 16 | the–moral–of–the | in–a–state–of | was–one–of–the | on–the–basis–of |
| 17 | the–business–of–the | for–the–purpose–of | the–rest–of–the | for–the–first–time |
| 18 | so–much–as–a | in–the–way–of | the–head–of–the | did–not–want–to |
| 19 | reader–may–please–to | the–hands–of–the | in–the–course–of | at–the–same–time |
| 20 | me–in–mind–of | on–the–other–hand | the–part–of–the | on–the–other–hand |
| 21 | it–may–be–so | for–the–most–part | the–middle–of–the | i–do–not–think |
| 22 | in–the–time–of | for–the–sake–of | a–member–of–the | in–the–middle–of |
| 23 | in–the–reign–of | the–middle–of–the | on–the–other–hand | the–top–of–the |
| 24 | i–shall–endeavour–to | the–town–of–mansoul | at–the–time–of | at–the–time–of |
| 25 | he–is–pleased–to | is–one–of–the | at–the–head–of | can–be–used–to |
| 26 | far–as–it–appears | at–the–end–of | in–the–form–of | the–middle–of–the |
| 27 | end–of–the–play | the–part–of–the | a–part–of–the | in–front–of–the |
| 28 | but–i–'–m | into–the–hands–of | for–the–sake–of | i–do–not–want |
| 29 | at–the–end–of | for–the–first–time | the–name–of–the | the–beginning–of–the |
| 30 | as–well–as–the | the–nature–of–the | in–the–hands–of | at–the–university–of |

**Table S4.** Top English 4-grams for arbitrary years.

| Rank | 1700 | 1800 | 1900 | 2000 |
|---|---|---|---|---|
| 1 | and–this–may–serve–to | on–the–part–of–the | on–the–part–of–the | at–the–end–of–the |
| 2 | as–far–as–it–appears | in–the–name–of–the | at–the–end–of–the | in–the–middle–of–the |
| 3 | will–not–so–much–as | and–at–the–same–time | and–at–the–same–time | i–do–not–want–to |
| 4 | we–have–no–reason–to | at–the–head–of–the | at–the–time–of–the | the–united–states–of–america |
| 5 | there–'s–no–need–of | in–the–midst–of–the | the–other–side–of–the | i–do–not–know–what |
| 6 | the–root–of–all–evil | the–other–side–of–the | at–the–head–of–the | the–other–side–of–the |
| 7 | the–heat–of–the–climate | into–the–hands–of–the | in–the–middle–of–the | at–the–time–of–the |
| 8 | the–end–of–the–play | the–name–of–the–lord | in–the–case–of–the | at–the–beginning–of–the |
| 9 | the–ancient–and–modern–stages | at–the–end–of–the | is–one–of–the–most | as–a–result–of–the |
| 10 | puts–me–in–mind–of | in–the–middle–of–the | in–the–hands–of–the | on–the–part–of–the |
| 11 | not–so–much–as–offer | in–the–hands–of–the | in–the–midst–of–the | is–one–of–the–most |
| 12 | not–so–much–as–a | on–the–banks–of–the | in–the–form–of–a | in–the–united–states–of |
| 13 | may–not–be–amiss–to | at–the–head–of–the | on–the–other–side–of | at–the–top–of–the |
| 14 | it–not–been–for–the | in–the–course–of–the | at–the–foot–of–the | at–the–end–of–the |
| 15 | it–may–not–be–amiss | on–the–other–side–of | in–the–direction–of–the | i–did–not–want–to |
| 16 | is–there–no–difference–between | of–the–town–of–mansoul | as–in–the–case–of | in–the–form–of–a |
| 17 | is–not–so–much–as | the–greater–part–of–the | to–be–found–in–the | at–the–bottom–of–the |
| 18 | is–an–odd–way–of | at–the–foot–of–the | in–the–history–of–the | on–the–other–side–of |
| 19 | i–'–m–afraid–i | at–the–bottom–of–the | i–do–not–know–what | in–such–a–way–that |
| 20 | from–whence–it–will–follow | the–inferior–extremity–of–the | at–the–beginning–of–the | in–the–united–states–and |
| 21 | does–not–so–much–as | in–the–beginning–of–the | in–the–course–of–the | for–the–first–time–in |
| 22 | does–not–in–the–least | at–the–head–of–his | the–greater–part–of–the | and–at–the–same–time |
| 23 | but–i–'–m–afraid | i–do–not–know–what | president–of–the–united–states | will–not–be–able–to |
| 24 | at–the–latter–end–of | the–inner–face–of–the | at–the–close–of–the | in–the–center–of–the |
| 25 | at–the–end–of–the | to–be–found–in–the | in–the–name–of–the | in–the–case–of–the |
| 26 | at–that–time–of–day | is–one–of–the–most | into–the–hands–of–the | i–do–not–know–how |
| 27 | as–it–would–be–to | president–of–the–united–states | at–the–bottom–of–the | printed–in–the–united–states |
| 28 | ancient–and–modern–stages–surveyed | in–the–form–of–a | i–do–not–want–to | by–the–end–of–the |
| 29 | according–to–the–custom–of | the–posterior–face–of–the | was–a–member–of–the | you–do–not–have–to |
| 30 | a–great–part–of–the | the–superior–extremity–of–the | on–the–side–of–the | you–do–not–have–to |

**Table S5.** Top English 5-grams for arbitrary years.

| Rank | 1700 | 1800 | 1900 | 2000 |
|------|------|------|------|------|
| 1 | de | de | de | de |
| 2 | la | la | la | la |
| 3 | que | et | et | et |
| 4 | le | les | à | à |
| 5 | les | le | les | le |
| 6 | à | à | le | des |
| 7 | qui | des | des | les |
| 8 | en | que | en | en |
| 9 | des | qui | du | du |
| 10 | du | dans | que | un |
| 11 | par | en | un | une |
| 12 | pour | du | une | que |
| 13 | dans | un | dans | dans |
| 14 | ne | une | qui | qui |
| 15 | ce | par | par | par |
| 16 | est | est | est | est |
| 17 | un | plus | pour | pour |
| 18 | il | ne | il | au |
| 19 | a | il | au | pas |
| 20 | vous | pour | a | sur |
| 21 | une | se | ne | a |
| 22 | l | ce | plus | plus |
| 23 | nous | a | se | se |
| 24 | plus | sur | pas | ne |
| 25 | au | on | sur | il |
| 26 | pas | au | ce | ce |
| 27 | se | pas | ou | ou |
| 28 | d | ou | on | la |
| 29 | on | l | nous | le |
| 30 | y | sont | sont | sont |

**Table S6.** Top French 1-grams for arbitrary years.

| Rank | **1700** | **1800** | **1900** | **2000** |
|---|---|---|---|---|
| 1 | de–la | de–la | de–la | de–la |
| 2 | à–la | à–la | à–la | à–la |
| 3 | que–les | et–de | et–de | et–de |
| 4 | dans–le | dans–les | dans–les | dans–le |
| 5 | que–le | que–les | dans–le | dans–la |
| 6 | dans–la | dans–le | dans–la | dans–les |
| 7 | que–la | dans–la | que–les | et–les |
| 8 | de–l | et–les | que–le | et–la |
| 9 | ce–qui | que–le | et–les | que–les |
| 10 | que–nous | que–la | que–la | et–le |
| 11 | tous–les | tous–les | et–la | que–le |
| 12 | que–vous | et–la | et–le | sur–le |
| 13 | dans–les | de–ces | par–le | que–la |
| 14 | ce–que | et–le | sur–les | de–l |
| 15 | de–ce | de–l | par–la | sur–la |
| 16 | qui–est | de–ce | sur–la | par–le |
| 17 | de–dieu | de–cette | tous–les | et–des |
| 18 | par–le | sur–les | par–les | par–la |
| 19 | par–la | par–la | sur–le | ce–qui |
| 20 | il–est | par–le | de–ces | sur–les |
| 21 | toutes–les | toutes–les | de–ce | par–les |
| 22 | ceux–qui | ce–qui | et–des | de–son |
| 23 | de–reims | sur–la | a–été | de–ce |
| 24 | y–a | par–les | de–son | à–une |
| 25 | de–cette | de–son | de–cette | pour–les |
| 26 | qui–ne | de–ses | de–l | de–cette |
| 27 | par–les | sur–le | ce–qui | et–à |
| 28 | que–de | et–des | pour–les | à–un |
| 29 | de–ladite | que–nous | pour–la | dans–un |
| 30 | de–ces | de–leur | y–a | de–ces |

**Table S7.** Top French 2-grams for arbitrary years.

| Rank | **1700** | **1800** | **1900** | **2000** |
|---|---|---|---|---|
| 1 | de–tous–les | et–de–la | et–de–la | et–de–la |
| 2 | mil–six–cens | de–tous–les | il–y–a | à–la–fois |
| 3 | il–y–a | plus–ou–moins | point–de–vue | il–y–a |
| 4 | tout–ce–qui | il–y–a | de–la–loi | ne–sont–pas |
| 5 | que–nous–avons | de–toutes–les | au–point–de | de–la–vie |
| 6 | ce–qui–est | le–nom–de | de–la–société | point–de–vue |
| 7 | de–ceux–qui | de–la–nature | plus–ou–moins | à–partir–de |
| 8 | tout–ce–que | de–la–terre | ne–sont–pas | plus–en–plus |
| 9 | duc–de–reims | ne–sont–pas | en–même–temps | la–mise–en |
| 10 | de–toutes–les | tout–ce–qui | il–y–a | de–plus–en |
| 11 | six–cens–quatre | il–y–a | de–tous–les | dans–le–cadre |
| 12 | la–somme–des | de–la–même | à–peu–près | ce–qui–est |
| 13 | ce–que–vous | la–plus–grande | la–loi–du | et–à–la |
| 14 | ce–que–nous | que–nous–avons | que–nous–avons | de–la–société |
| 15 | tous–les–jours | grand–nombre–de | à–la–fois | de–la–population |
| 16 | la–somme–de | dans–tous–les | le–nom–de | il–y–a |
| 17 | ville–de–reims | un–grand–nombre | de–la–vie | par–rapport–à |
| 18 | de–la–nature | de–la–mer | et–à–la | en–tant–que |
| 19 | pieces–de–comparaison | ce–qui–est | ce–qui–concerne | à–la–fin |
| 20 | de–ce–que | et–à–la | de–la–ville | la–fin–de |
| 21 | archevêque–duc–de | de–ceux–qui | le–droit–de | la–fin–du |
| 22 | entre–les–mains | une–espèce–de | de–toutes–les | de–la–ville |
| 23 | amour–de–dieu | dans–toutes–les | de–la–france | plus–ou–moins |
| 24 | à–ceux–qui | la–plupart–des | tout–à–fait | de–la–france |
| 25 | du–mois–de | à–tous–les | ce–qui–est | en–même–temps |
| 26 | aprés–le–choc | que–dans–les | de–plus–en | de–tous–les |
| 27 | que–ce–soit | au–lieu–de | plus–en–plus | de–la–loi |
| 28 | que–vous–ne | de–la–vie | chemins–de–fer | la–plupart–des |
| 29 | que–nous–ne | de–la–france | chemin–de–fer | de–la–république |
| 30 | il–est–évident | on–ne–peut | la–fin–de | ce–qui–concerne |

**Table S8.** Top French 3-grams for arbitrary years.

| Rank | 1700 | 1800 | 1900 | 2000 |
|------|------|------|------|------|
| 1 | mil–six–cens–quatre | un–grand–nombre–de | au–point–de–vue | de–plus–en–plus |
| 2 | archevêque–duc–de–reims | sous–le–nom–de | de–plus–en–plus | dans–la–mesure–où |
| 3 | premier–pair–de–france | il–y–en–a | point–de–vue–de | dans–le–cadre–de |
| 4 | mil–six–cens–soixante | la–plus–grande–partie | de–la–loi–du | en–ce–qui–concerne |
| 5 | il–est–évident–que | de–la–même–manière | en–ce–qui–concerne | un–certain–nombre–de |
| 6 | la–grace–de–dieu | de–plus–en–plus | sous–le–nom–de | la–mise–en–place |
| 7 | procureur–general–du–roy | de–la–loi–du | un–certain–nombre–de | du–point–de–vue |
| 8 | la–commodité–ou–incommodité | que–je–viens–de | un–grand–nombre–de | à–la–fin–du |
| 9 | à–la–requeste–de | la–surface–de–la | en–même–temps–que | pour–la–première–fois |
| 10 | de–ladite–eglise–de | que–nous–venons–de | des–chemins–de–fer | à–la–fin–de |
| 11 | tout–ce–que–vous | il–y–a–des | de–vue–de–la | la–mise–en–œuvre |
| 12 | par–la–grace–de | en–est–de–même | à–la–fin–de | la–fin–de–la |
| 13 | de–tout–ce–qui | à–la–fin–de | ce–qui–concerne–les | au–sein–de–la |
| 14 | des–pieces–de–comparaison | de–tout–ce–qui | pour–la–première–fois | dans–le–domaine–de |
| 15 | la–gloire–de–dieu | qui–ne–sont–pas | du–chemin–de–fer | par–rapport–à–la |
| 16 | il–est–aisé–de | toutes–les–parties–de | à–la–suite–de | en–ce–qui–concerne |
| 17 | de–la–ville–de | par–le–moyen–de | de–la–ville–de | sous–la–direction–de |
| 18 | commodité–ou–incommodité–de | de–la–république–française | en–ce–qui–concerne | ce–qui–concerne–les |
| 19 | la–piece–arguée–de | les–uns–des–autres | de–la–loi–de | la–fin–des–années |
| 20 | de–ladite–abbaye–de | une–grande–quantité–de | ce–point–de–vue | la–question–de–la |
| 21 | tout–ce–qui–est | à–la–tête–de | en–est–de–même | un–point–de–vue |
| 22 | il–est–certain–que | pour–la–première–fois | ce–qui–concerne–la | point–de–vue–de |
| 23 | aprés–la–mort–de | il–y–a–des | il–ne–faut–pas | le–cadre–de–la |
| 24 | par–le–moyen–de | plus–grande–partie–de | que–nous–venons–de | au–cours–de–la |
| 25 | il–ne–faut–pas | il–ne–faut–pas | la–plus–grande–partie | à–partir–de–la |
| 26 | de–nostredite–ville–de | de–la–part–de | y–a–lieu–de | de–la–mise–en |
| 27 | contre–les–conclusions–du | de–la–plus–grande | il–y–a–des | de–la–part–de |
| 28 | à–la–somme–des | le–plus–grand–nombre | de–la–part–de | à–la–suite–de |
| 29 | les–conclusions–du–demandeur | de–la–part–des | de–la–part–de | de–la–part–des |
| 30 | jour–du–mois–de | dans–le–cours–de | de–chemins–de–fer | qui–ne–sont–pas |

**Table S9.** Top French 4-grams for arbitrary years.

| Rank | 1700 | 1800 | 1900 | 2000 |
|------|------|------|------|------|
| 1 | par–la–grace–de–dieu | la–plus–grande–partie–de | au–point–de–vue–de | dans–le–cadre–de–la |
| 2 | la–commodité–ou–incommodité–de | il–en–est–de–même | point–de–vue–de–la | dans–le–domaine–de–la |
| 3 | pour–y–avoir–recours–en | de–la–manière–la–plus | en–ce–qui–concerne–les | du–point–de–vue–de |
| 4 | la–grace–de–dieu–roy | suit–la–teneur–de–la | à–ce–point–de–vue | au–fur–et–à–mesure |
| 5 | grace–de–dieu–roy–de | la–teneur–de–la–déclaration | au–fur–et–à–mesure | en–ce–qui–concerne–les |
| 6 | de–dieu–roy–de–france | dont–nous–venons–de–parler | il–en–est–de–même | la–mise–en–place–de |
| 7 | cour–de–parlement–de–paris | la–surface–de–la–terre | au–point–de–vue–du | la–mise–en–œuvre–de |
| 8 | de–grace–mil–six–cens | anciens–approuve–la–résolution–ci | en–ce–qui–concerne–la | en–ce–qui–concerne–la |
| 9 | louis–par–la–grace–de | dont–je–viens–de–parler | au–point–de–vue–des | à–la–fin–des–années |
| 10 | par–ces–presentes–signées–de | la–présente–résolution–sera–imprimée | il–y–a–lieu–de | dans–la–mesure–où–il |
| 11 | de–la–commodité–ou–incommodité | conseil–des–anciens–approuve–la | la–plus–grande–partie–de | point–de–vue–de–la |
| 12 | y–avoir–recours–en–cas | des–anciens–approuve–la–résolution | il–y–a–quelques–années | à–la–fin–de–la |
| 13 | du–grand–sceau–de–cire | urgence–et–de–la–résolution | en–ce–qui–concerne–le | en–ce–qui–concerne–les |
| 14 | car–tel–est–nostre–plaisir | qui–précède–la–résolution–ci | en–ce–qui–concerne–les | de–ce–point–de–vue |
| 15 | avoir–recours–en–cas–de | le–conseil–des–anciens–approuve | du–chemin–de–fer–de | dans–la–mesure–où–elle |
| 16 | ordonné–ce–que–de–raison | urgence–qui–précède–la–résolution | à–la–suite–de–la | sur–le–marché–du–travail |
| 17 | les–lettres–patentes–du–roy | et–de–la–résolution–du | des–chemins–de–fer–de | de–ce–point–de–vue |
| 18 | la–commodité–ou–incommodité–que | inséré–au–bulletin–des–lois | il–en–est–de–même | réduction–du–temps–de–travail |
| 19 | de–la–cour–de–parlement | nom–de–la–republique–française | à–la–fin–de–la | la–fin–du–xixe–siècle |
| 20 | égale–à–la–somme–des | au–nom–de–la–republique | au–point–de–vue–de | fur–et–à–mesure–que |
| 21 | patentes–du–roy–données–à | adoptant–les–motifs–de–la | de–la–façon–la–plus | en–ce–qui–concerne–le |
| 22 | le–procureur–general–du–roy | sera–inséré–au–bulletin–des | dans–le–sens–de–la | la–mise–en–œuvre–des |
| 23 | du–procureur–general–du–roy | un–plus–grand–nombre–de | la–séance–est–levée–à | de–plus–en–plus–de |
| 24 | pour–en–jouir–par–ledit | toutes–les–parties–de–la | la–question–de–savoir–si | la–fin–de–la–guerre |
| 25 | la–chambre–des–comptes–de | les–unes–sur–les–autres | dans–la–plupart–des–cas | en–ce–qui–concerne–la |
| 26 | grand–sceau–de–cire–verte | de–la–même–manière–que | le–président–de–la–république | dans–la–mesure–où–les |
| 27 | de–monsieur–le–procureur–general | en–ce–qui–concerne | en–ce–qui–concerne–la | à–la–suite–de–la |
| 28 | chambre–des–comptes–de–paris | sur–le–rapport–du–ministre | en–même–temps–que–la | est–de–plus–en–plus |
| 29 | recours–en–cas–de–besoin | les–uns–sur–les–autres | en–même–temps–que–le | il–en–est–de–même |
| 30 | quatre–mille–cinq–cens–livres | qui–sera–inséré–au–bulletin | de–la–manière–la–plus | de–la–mise–en–œuvre |

**Table S10.** Top French 5-grams for arbitrary years.

| Rank | **1700** | **1800** | **1900** | **2000** |
|------|----------|----------|----------|----------|
| 1 | und | der | der | der |
| 2 | die | und | die | die |
| 3 | in | die | und | und |
| 4 | zu | in | in | in |
| 5 | der | zu | den | von |
| 6 | daß | den | des | den |
| 7 | von | von | zu | zu |
| 8 | ist | des | von | des |
| 9 | erfahrung | ist | ist | das |
| 10 | nicht | nicht | das | sich |
| 11 | sie | das | sich | ist |
| 12 | den | dem | nicht | mit |
| 13 | sich | so | dem | auf |
| 14 | mit | sich | mit | nicht |
| 15 | ein | sie | auf | im |
| 16 | dem | als | eine | für |
| 17 | als | mit | als | als |
| 18 | des | auf | im | eine |
| 19 | werden | ein | auch | die |
| 20 | welche | es | ein | dem |
| 21 | so | oder | die | auch |
| 22 | durch | eine | für | ein |
| 23 | bey | daß | an | an |
| 24 | das | er | so | werden |
| 25 | wird | auch | es | es |
| 26 | man | werden | sie | sie |
| 27 | sind | durch | durch | einer |
| 28 | sey | man | bei | daß |
| 29 | können | an | dass | aus |
| 30 | ich | aber | nach | oder |

**Table S11.** Top German 1-grams for arbitrary years.

| Rank | **1700** | **1800** | **1900** | **2000** |
|---|---|---|---|---|
| 1 | die–erfahrung | in–der | in–der | in–der |
| 2 | daß–sie | und–die | in–den | für–die |
| 3 | durch–die | in–den | für–die | in–den |
| 4 | von–der | von–der | und–die | und–die |
| 5 | in–der | und–der | von–der | auf–die |
| 6 | hohen–alters | auf–die | auf–die | und–der |
| 7 | des–hohen | durch–die | durch–die | mit–der |
| 8 | zierde–des | von–dem | und–der | von–der |
| 9 | und–in | in–dem | mit–der | mit–dem |
| 10 | sich–in | mit–dem | mit–dem | in–die |
| 11 | in–denselben | von–den | über–die | über–die |
| 12 | in–den | mit–der | in–die | durch–die |
| 13 | in–dem | daß–die | dass–die | daß–die |
| 14 | hohen–alter | in–die | bei–der | bei–der |
| 15 | der–gottesfurcht | für–die | in–dem | an–der |
| 16 | daß–ich | über–die | von–den | sich–die |
| 17 | von–den | auf–den | an–der | auch–die |
| 18 | von–dem | daß–sie | auch–die | für–den |
| 19 | theils–an | und–in | aus–dem | aus–der |
| 20 | sie–nicht | auch–die | auf–den | aus–dem |
| 21 | man–nicht | als–die | von–dem | auf–den |
| 22 | ist–es | eben–so | ist–die | auf–der |
| 23 | erfahrung–und | aus–dem | mit–den | vor–allem |
| 24 | der–erfahrung | daß–er | aus–der | mit–den |
| 25 | dem–hohen | in–einem | an–die | von–den |
| 26 | daß–er | aus–der | an–den | ist–die |
| 27 | an–andern | und–den | für–den | nicht–nur |
| 28 | zu–suchen | und–das | bei–den | an–die |
| 29 | zu–allen | wenn–sie | sich–die | in–einem |
| 30 | wird–ihm | so–wie | nach–der | die–sich |

**Table S12.** Top German 2-grams for arbitrary years.

| Rank | 1700 | 1800 | 1900 | 2000 |
|---|---|---|---|---|
| 1 | des–hohen–alters | so–ist–es | in–der–regel | in–der–regel |
| 2 | dem–hohen–alter | in–der–folge | in–bezug–auf | die–in–der |
| 3 | von–jugend–auf | nach–und–nach | und–in–der | und–in–der |
| 4 | sich–in–der | und–in–der | in–der–that | sich–in–der |
| 5 | größte–zierde–des | in–ansehung–der | mehr–oder–weniger | im–rahmen–der |
| 6 | einzig–und–allein | so–wie–die | es–sich–um | im–hinblick–auf |
| 7 | die–größte–zierde | in–der–natur | die–zahl–der | auch–in–der |
| 8 | zu–was–vor | in–der–that | sich–in–der | es–sich–um |
| 9 | wird–ihm–so | eben–so–wenig | die–in–der | in–erster–linie |
| 10 | wir–nach–und | in–so–fern | auch–in–der | in–der–ddr |
| 11 | wenn–wir–sie | in–der–welt | in–erster–linie | handelt–es–sich |
| 12 | von–der–wahrheit | mehr–oder–weniger | in–der–mitte | in–der–lage |
| 13 | und–in–dem | asthenie–der–erregung | in–den–letzten | im–bereich–der |
| 14 | theils–an–andern | der–andern–seite | bezug–auf–die | im–zusammenhang–mit |
| 15 | sie–sind–nicht | sich–in–der | in–diesem–falle | in–den–letzten |
| 16 | sie–nicht–vor | und–in–den | in–der–nähe | vor–allem–die |
| 17 | sich–auch–die | die–in–der | und–in–der | nicht–nur–die |
| 18 | nicht–nur–bey | in–der–mitte | auch–für–die | auch–für–die |
| 19 | nach–und–nach | auch–in–der | auf–grund–der | frankfurt–am–main |
| 20 | mit–recht–von | so–kann–man | sich–in–den | in–diesem–zusammenhang |
| 21 | mit–diesem–doppelten | auf–diese–art | handelt–es–sich | in–der–bundesrepublik |
| 22 | mehr–und–mehr | eine–art–von | so–ist–es | der–bundesrepublik–deutschland |
| 23 | kräfte–des–heiligen | daß–sie–sich | die–in–den | sich–in–den |
| 24 | in–welchen–sie | zu–gleicher–zeit | auch–in–den | sich–auf–die |
| 25 | in–der–jugend | der–natur–der | eine–reihe–von | nach–wie–vor |
| 26 | in–dem–verstande | und–es–ist | in–der–weise | die–frage–nach |
| 27 | erfahrung–ist–es | in–beziehung–auf | auf–diese–weise | die–in–den |
| 28 | durch–die–erfahrung | daß–er–sich | rücksicht–auf–die | hinblick–auf–die |
| 29 | die–sich–in | und–von–der | mehr–und–mehr | in–den–usa |
| 30 | die–erfahrung–ist | der–gewalt–des | der–in–der | und–vor–allem |

**Table S13.** Top German 3-grams for arbitrary years.

| Rank | 1700 | 1800 | 1900 | 2000 |
|---|---|---|---|---|
| 1 | die–größte–zierde–des | von–zeit–zu–zeit | in–bezug–auf–die | im–hinblick–auf–die |
| 2 | wir–nach–und–nach | auf–der–andern–seite | auf–dem–gebiete–der | handelt–es–sich–um |
| 3 | sich–in–der–jugend | der–natur–der–sache | an–und–für–sich | in–den–letzten–jahren |
| 4 | sich–auch–die–erfahrung | an–und–für–sich | in–den–letzten–jahren | auf–der–anderen–seite |
| 5 | kräfte–des–heiligen–geistes | verminderung–der–gewalt–des | mit–rücksicht–auf–die | in–der–bundesrepublik–deutschland |
| 6 | die–sich–in–der | so–viel–als–möglich | in–den–meisten–fällen | im–zusammenhang–mit–der |
| 7 | die–erfahrung–ist–es | in–der–mitte–des | in–der–nähe–der | auf–dem–gebiet–der |
| 8 | zum–nutzen–gereichen–können | auf–der–einen–seite | auf–der–anderen–seite | die–frage–nach–der |
| 9 | zu–welcher–wir–nach | die–art–und–weise | es–sich–um–die | in–bezug–auf–die |
| 10 | zu–was–vor–einem | in–der–natur–der | bis–zu–einem–gewissen | es–sich–um–eine |
| 11 | zu–thun–verbunden–sind | immer–mehr–und–mehr | von–zeit–zu–zeit | auf–der–anderen–seite |
| 12 | zu–dem–genüsse–der | der–könig–von–siam | es–sich–um–eine | in–den–neuen–bundesländern |
| 13 | zu–allen–zeiten–gegen | in–absicht–auf–die | wenn–es–sich–um | auf–den–ersten–blick |
| 14 | zu–allen–dingen–nütze | auf–der–andern–seite | auf–der–einen–seite | nicht–in–der–lage |
| 15 | zeiget–sich–auch–die | in–hinsicht–auf–die | handelt–es–sich–um | es–handelt–sich–um |
| 16 | wohl–nicht–auf–diese | in–so–fern–sie | an–die–stelle–der | in–bezug–auf–die |
| 17 | wird–nicht–mehr–von | auf–irgend–eine–art | an–die–stelle–der | stellt–sich–die–frage |
| 18 | wird–man–nicht–über | aus–der–natur–der | zu–einem–gewissen–grade | in–der–zweiten–hälfte |
| 19 | wird–ihm–nicht–zu | der–lehre–von–der | auf–der–ersten–blick | der–zweiten–hälfte–des |
| 20 | wird–die–erfahrung–durch | der–absoluten–gewalt–des | auf–dem–wege–der | auf–der–ebene–der |
| 21 | wir–werden–vielmehr–in | das–ding–an–sich | im–laufe–der–zeit | auf–der–einen–seite |
| 22 | wir–auch–den–werth | so–ist–es–auch | auf–dem–gebiete–des | vor–allem–in–der |
| 23 | wir–am–deutlichsten–aus | so–ist–es–doch | an–ort–und–stelle | die–frage–nach–dem |
| 24 | wie–sie–sich–in | die–dinge–an–sich | die–art–und–weise | nach–dem–zweiten–weltkrieg |
| 25 | wie–oft–haben–nicht | kritik–der–reinen–vernunft | nicht–die–rede–sein | vor–dem–hintergrund–der |
| 26 | wie–es–mit–den | die–lehre–von–der | in–der–mitte–des | im–zusammenhang–mit–dem |
| 27 | wie–dieses–oder–jenes | auf–den–heutigen–tag | die–frage–nach–der | im–vergleich–zu–den |
| 28 | werden–vielmehr–in–den | in–beziehung–auf–die | in–erster–linie–die | vor–allem–in–den |
| 29 | werden–dieselben–von–der | an–die–stelle–der | in–der–mitte–der | auf–der–grundlage–der |
| 30 | wenn–wir–sie–in | nur–unter–der–bedingung | in–bezug–auf–den | in–der–lage–ist |

**Table S14.** Top German 4-grams for arbitrary years.

| Rank | 1700 | 1800 | 1900 | 2000 |
|------|------|------|------|------|
| 1 | die–sich–in–der–jugend | bis–auf–den–heutigen–tag | bis–zu–einem–gewissen–grade | in–der–zweiten–hälfte–des |
| 2 | wird–nicht–mehr–von–den | aus–der–natur–der–sache | bis–auf–den–heutigen–tag | handelt–es–sich–um–eine |
| 3 | wird–man–nicht–über–die | wie–ich–schon–gesagt–habe | in–der–zweiten–hälfte–des | in–der–ersten–hälfte–des |
| 4 | wir–nach–und–nach–zu | an–und–für–sich–selbst | in–der–mehrzahl–der–fälle | in–der–zweiten–hälfte–der |
| 5 | wir–nach–und–nach–durch | über–das–verhalten–des–phosphors | in–der–ersten–hälfte–des | kölner–zeitschrift–für–soziologie–und |
| 6 | wir–auch–den–werth–der | versuche–über–das–verhalten–des | der–kaiserlichen–akademie–der–wissenschaften | zeitschrift–für–soziologie–und–sozialpsychologie |
| 7 | wir–am–deutlichsten–aus–den | der–natur–der–sache–nach | in–der–natur–der–sache | es–handelt–sich–dabei–um |
| 8 | wie–sie–sich–in–allen | in–der–natur–der–sache | zum–wirklichen–mitgliede–ernannt–am | handelt–es–sich–um–die |
| 9 | wie–es–mit–den–menschen | des–raums–und–der–zeit | die–eine–oder–die–andere | auf–der–einen–seite–und |
| 10 | wenn–sie–sich–auf–den | des–menschen–gegen–sich–selbst | an–der–universität–in–wien | o–o–o–o–o |
| 11 | wenn–er–von–der–wahrheit | die–eine–oder–die–andere | dass–es–sich–hier–um | handelt–es–sich–um–einen |
| 12 | sie–sind–nicht–nur–zu | versteht–es–sich–von–selbst | wenn–es–sich–um–die | handelt–es–sich–um–eine |
| 13 | sich–nach–dem–lauf–der | und–auf–der–andern–seite | dass–es–sich–um–eine | dabei–handelt–es–sich–um |
| 14 | sich–einzig–und–allein–zu | aus–dem–wege–zu–räumen | nicht–die–rede–sein–kann | es–handelt–sich–um–eine |
| 15 | sich–ihr–ganzes–leben–hindurch | a–n–m–e–r | der–natur–der–sache–nach | stellt–sich–die–frage–nach |
| 16 | noch–in–der–blüthe–ihrer | religion–innerhalb–der–grenzen–der | von–der–hand–zu–weisen | auch–im–hinblick–auf–die |
| 17 | nicht–auf–diese–weise–die | in–den–stand–zu–setzen | bis–in–die–neueste–zeit | sich–in–den–letzten–jahren |
| 18 | nach–und–nach–zu–der | geschichte–der–lehre–von–der | handelt–es–sich–um–eine | das–gilt–insbesondere–für–vervielfältigungen |
| 19 | nach–und–nach–durch–die | das–moralische–gesetz–in–seine | die–art–und–weise–der | mikroverfilmungen–und–die–einspeicherung–und |
| 20 | nach–dem–lauf–der–natur | in–der–kasse–der–bank | den–vereinigten–staaten–von–amerika | hierbei–handelt–es–sich–um |
| 21 | mit–recht–von–sich–rühmen | grundlegung–zur–metaphysik–der–sitten | der–einen–oder–der–anderen | bis–zu–einem–gewissen–grad |
| 22 | man–sich–in–seinem–eigenen | des–angriffs–und–der–vertheidigung | handelt–es–sich–um–die | und–verarbeitung–in–elektronischen–systemen |
| 23 | in–der–heiligen–schrift–vor | nicht–mehr–und–nicht–weniger | reich–gottes–ist–in–euch | einspeicherung–und–verarbeitung–in–elektronischen |
| 24 | in–der–blüthe–ihrer–jahre | liegt–in–der–natur–der | das–reich–gottes–ist–in | und–die–einspeicherung–und–verarbeitung |
| 25 | in–den–gemüthern–der–menschen | in–der–ersten–hälfte–des | schon–an–und–für–sich | grenzen–des–urheberrechtsgesetzes–ist–ohne |
| 26 | ich–nicht–zu–viel–sage | daseyn–gottes–und–die–unsterblichkeit | es–liegt–auf–der–hand | die–einspeicherung–und–verarbeitung–in |
| 27 | es–ist–nicht–zu–leugnen | wenn–sie–auch–noch–so | es–handelt–sich–hier–um | mehr–als–die–hälfte–der |
| 28 | es–ist–also–leicht–zu | so–viel–als–möglich–zu | ist–hier–nicht–der–ort | im–wahrsten–sinne–des–wortes |
| 29 | einzig–und–allein–aus–der | raum–und–in–der–zeit | wo–es–sich–um–die | in–erster–linie–auf–die |
| 30 | einer–mehr–als–der–andere | moralische–gesetz–in–seine–maxime | verhält–es–sich–mit–der | nicht–mehr–in–der–lage |

**Table S15.** Top German 5-grams for arbitrary years.

| Rank | 1700 | 1800 | 1900 | 2000 |
|------|------|------|------|------|
| 1 | di | di | di | di |
| 2 | del | e | e | e |
| 3 | e | che | che | che |
| 4 | in | la | la | la |
| 5 | la | il | il | in |
| 6 | della | in | in | il |
| 7 | il | a | del | a |
| 8 | che | per | a | del |
| 9 | per | non | per | della |
| 10 | non | si | non | per |
| 11 | a | del | si | un |
| 12 | n | le | della | è |
| 13 | al | i | un | non |
| 14 | con | della | le | si |
| 15 | un | un | è | una |
| 16 | alla | da | i | le |
| 17 | da | ed | una | con |
| 18 | rei | con | da | i |
| 19 | c | più | dei | da |
| 20 | ed | una | con | dei |
| 21 | cui | de | al | al |
| 22 | è | al | più | nel |
| 23 | sez | è | delle | delle |
| 24 | civ | delle | nel | come |
| 25 | dei | o | ed | più |
| 26 | una | nel | alla | alla |
| 27 | o | alla | o | o |
| 28 | nel | gli | come | la |
| 29 | delle | l | gli | anche |
| 30 | le | se | nella | nella |

**Table S16.** Top Italian 1-grams for arbitrary years.

Supplementary Material| Rank | **1700** | **1800** | **1900** | **2000** |
|---|---|---|---|---|
| 1 | di–un | e–di | e–di | di–un |
| 2 | di–cui | che–si | per–la | di–una |
| 3 | in–cui | e–la | che–si | e–di |
| 4 | ai–sensi | che–non | che–il | per–la |
| 5 | con–rinvio | per–la | e–la | e–la |
| 6 | cassa–con | e–che | che–la | che–si |
| 7 | per–la | che–il | di–un | con–la |
| 8 | di–lavoro | che–la | che–non | e–il |
| 9 | di–una | non–si | non–si | in–cui |
| 10 | del–giudice | e–per | non–è | che–la |
| 11 | da–parte | di–un | e–che | per–il |
| 12 | per–il | tutte–le | di–una | la–sua |
| 13 | e–non | e–le | la–sua | che–il |
| 14 | ex–art | la–sua | e–il | che–non |
| 15 | con–la | il–quale | in–cui | non–è |
| 16 | non–è | che–in | e–non | in–un |
| 17 | che–il | ed–il | e–le | con–il |
| 18 | tema–di | tutti–i | in–un | il–suo |
| 19 | non–può | che–i | il–suo | e–le |
| 20 | nei–confronti | di–questa | tutte–le | non–si |
| 21 | che–la | di–questo | per–le | di–cui |
| 22 | di–merito | che–le | e–per | si–è |
| 23 | caso–di | la–quale | di–cui | e–che |
| 24 | in–sede | e–non | con–la | in–una |
| 25 | ed–altro | in–un | il–quale | e–non |
| 26 | ed–altri | che–per | per–il | da–un |
| 27 | a–norma | i–quali | tutti–i | che–è |
| 28 | secondo–comma | e–si | e–si | e–della |
| 29 | giudice–di | di–lui | la–quale | e–i |
| 30 | in–tema | e–con | si–è | per–le |

**Table S17.** Top Italian 2-grams for arbitrary years.

| Rank | **1700** | **1800** | **1900** | **2000** |
|---|---|---|---|---|
| 1 | cassa–con–rinvio | che–non–si | più–o–meno | a–cura–di |
| 2 | in–tema–di | di–tutte–le | non–si–può | punto–di–vista |
| 3 | caso–in–cui | di–tutti–i | che–non–si | in–grado–di |
| 4 | in–sede–di | il–nome–di | di–tutte–le | una–serie–di |
| 5 | da–parte–del | per–lo–più | in–cui–si | in–cui–si |
| 6 | competenza–e–giurisdizione | tutti–gli–altri | il–diritto–di | una–sorta–di |
| 7 | datore–di–lavoro | e–per–la | di–tutti–i | la–possibilità–di |
| 8 | giudice–di–merito | in–tutte–le | e–per–la | il–fatto–che |
| 9 | la–corte–suprema | tutto–ciò–che | il–nome–di | da–parte–di |
| 10 | ne–consegue–che | poco–a–poco | disegno–di–legge | si–tratta–di |
| 11 | nel–caso–in | più–o–meno | non–può–essere | più–o–meno |
| 12 | con–la–conseguenza | e–che–non | per–la–sua | la–prima–volta |
| 13 | imposte–e–tasse | vale–a–dire | punto–di–vista | in–materia–di |
| 14 | del–giudice–di | di–quello–che | la–prima–volta | che–non–si |
| 15 | ricorso–per–cassazione | se–non–che | che–non–è | e–la–sua |
| 16 | corte–suprema–ha | la–maggior–parte | in–tutte–le | al–di–là |
| 17 | la–conseguenza–che | in–cui–si | in–tutti–i | da–un–lato |
| 18 | in–cui–il | non–si–può | si–può–dire | in–questo–caso |
| 19 | in–caso–di | in–tutti–i | in–cui–la | dal–punto–di |
| 20 | del–datore–di | a–poco–a | consiglio–di–stato | in–cui–il |
| 21 | per–pubblica–utilità | la–qual–cosa | si–tratta–di | di–tutte–le |
| 22 | ai–fini–della | per–la–sua | in–cui–il | da–parte–del |
| 23 | cassa–senza–rinvio | e–gli–altri | una–specie–di | in–cui–la |
| 24 | nel–caso–in | il–duca–di | e–la–sua | al–fine–di |
| 25 | rapporto–di–lavoro | di–tutto–il | in–modo–che | per–la–prima |
| 26 | nel–caso–di | di–cui–si | tutti–gli–altri | si–tratta–di |
| 27 | espropriazione–per–pubblica | la–prima–volta | in–gran–parte | di–tutti–i |
| 28 | in–tema–di | di–tutti–gli | tutto–ciò–che | per–la–sua |
| 29 | nei–confronti–del | che–non–è | di–cui–si | non–può–essere |
| 30 | al–fine–di | per–mezzo–di | una–serie–di | la–presenza–di |

**Table S18.** Top Italian 3-grams for arbitrary years.

| Rank | 1700 | 1800 | 1900 | 2000 |
|---|---|---|---|---|
| 1 | nel–caso–in–cui | a–poco–a–poco | per–la–prima–volta | dal–punto–di–vista |
| 2 | con–la–conseguenza–che | c–sol–fa–ut | a–poco–a–poco | per–la–prima–volta |
| 3 | la–corte–suprema–ha | il–più–delle–volte | dal–punto–di–vista | nel–momento–in–cui |
| 4 | del–datore–di–lavoro | sotto–il–nome–di | del–consiglio–di–stato | una–vera–e–propria |
| 5 | del–giudice–di–merito | commissione–straordinaria–di–governo | si–può–dire–che | un–vero–e–proprio |
| 6 | espropriazione–per–pubblica–utilità | il–duca–di–savoja | sotto–il–nome–di | un–punto–di–vista |
| 7 | nel–caso–in–cui | g–sol–re–ut | ha–facoltà–di–parlare | di–volta–in–volta |
| 8 | in–sede–di–legittimità | di–mano–in–mano | di–grazia–e–giustizia | è–in–grado–di |
| 9 | specie–la–corte–suprema | per–lo–spazio–di | di–previsione–della–spesa | di–cui–al–comma |
| 10 | nella–specie–la–corte | per–la–qual–cosa | della–spesa–del–ministero | nel–caso–in–cui |
| 11 | considerazione–del–fatto–che | per–la–prima–volta | del–disegno–di–legge | questo–punto–di–vista |
| 12 | caso–in–cui–il | in–tutta–la–sua | previsione–della–spesa–del | si–tratta–di–un |
| 13 | in–considerazione–del–fatto | a–la–mi–re | un–gran–numero–di | a–che–fare–con |
| 14 | corte–suprema–ha–confermato | come–si–è–detto | stato–di–previsione–della | si–tratta–di–un |
| 15 | da–parte–del–giudice | di–giorno–in–giorno | in–questi–ultimi–anni | da–un–punto–di |
| 16 | ha–confermato–la–sentenza | di–tutti–gli–altri | un–certo–numero–di | al–di–là–di |
| 17 | provvedimenti–impugnabili–o–no | d–la–sol–re | dello–stato–di–previsione | nella–misura–in–cui |
| 18 | suprema–ha–confermato–la | per–la–qual–cosa | il–disegno–di–legge | dal–punto–di–vista |
| 19 | di–cui–agli–artt | il–capo–del–femore | ministero–dei–lavori–pubblici | si–tratta–di–una |
| 20 | per–la–prima–volta | per–la–maggior–parte | il–più–delle–volte | nella–seconda–metà–del |
| 21 | infortuni–sul–lavoro–e | di–tempo–in–tempo | nel–caso–in–cui | in–un–certo–senso |
| 22 | sul–lavoro–e–malattie | di–tutte–le–cose | in–tutta–la–sua | come–si–è–detto |
| 23 | lavoro–e–malattie–professionali | a–un–di–presso | per–ciò–che–riguarda | si–tratta–di–una |
| 24 | onere–di–prova–del | di–tutto–ciò–che | che–non–si–può | del–presidente–della–repubblica |
| 25 | opposizione–agli–atti–esecutivi | di–cui–si–tratta | legge–comunale–e–provinciale | come–si–è–visto |
| 26 | nella–parte–in–cui | del–fondo–di–religione | una–vera–e–propria | di–una–serie–di |
| 27 | ai–sensi–degli–artt | del–re–di–francia | segretario–di–stato–per | al–di–là–della |
| 28 | sentenza–della–corte–costituzionale | in–uno–stato–di | in–tutti–i–casi | il–modo–in–cui |
| 29 | il–giudice–di–merito | in–tutte–le–sue | ministero–dei–lavori–pubblici | si–può–dire–che |
| 30 | il–ricorso–per–cassazione | la–repubblica–di–venezia | che–si–tratti–di | per–quanto–riguarda–la |

**Table S19.** Top Italian 4-grams for arbitrary years.

| Rank | 1700 | 1800 | 1900 | 2000 |
|---|---|---|---|---|
| 1 | nella–specie–la–corte–suprema | in–tutta–la–sua–estensione | previsione–della–spesa–del–ministero | da–un–punto–di–vista |
| 2 | in–considerazione–del–fatto–che | tutto–il–territorio–della–repubblica | di–previsione–della–spesa–del | decreto–del–presidente–della–repubblica |
| 3 | specie–la–corte–suprema–ha | primo–console–della–repubblica–francese | stato–di–previsione–della–spesa | da–questo–punto–di–vista |
| 4 | la–corte–suprema–ha–confermato | munita–del–sigillo–della–repubblica | dello–stato–di–previsione–della | dal–punto–di–vista–della |
| 5 | corte–suprema–ha–confermato–la | di–nuovo–e–da–capo | disegni–di–legge–e–relazioni | data–di–entrata–in–vigore |
| 6 | sul–lavoro–e–malattie–professionali | in–tutte–le–sue–parti | cassa–dei–depositi–e–prestiti | da–questo–punto–di–vista |
| 7 | infortuni–sul–lavoro–e–malattie | in–tutto–il–territorio–della | della–legge–comunale–e–provinciale | nella–maggior–parte–dei–casi |
| 8 | nel–caso–in–cui–il | diritto–di–esercitare–la–professione | sezione–del–consiglio–di–stato | presidente–del–consiglio–dei–ministri |
| 9 | suprema–ha–confermato–la–sentenza | entro–il–termine–di–giorni | in–tutto–o–in–parte | per–la–prima–volta–in |
| 10 | sentenza–della–corte–costituzionale–n | ed–il–duca–di–savoja | delle–leggi–e–dei–decreti | non–è–in–grado–di |
| 11 | per–la–prima–volta–in | che–il–più–delle–volte | ufficiale–delle–leggi–e–dei | di–una–vera–e–propria |
| 12 | ricorso–per–cassazione–ai–sensi | il–diritto–di–esercitare–la | ministro–segretario–di–stato–per | del–presidente–del–consiglio–dei |
| 13 | ha–confermato–la–sentenza–impugnata | tutte–le–parti–del–corpo | leggi–e–dei–decreti–del | dal–punto–di–vista–del |
| 14 | la–sentenza–di–merito–che | della–repubblica–francese–una–ed | mandando–a–chiunque–spetti–di | sia–dal–punto–di–vista |
| 15 | nella–parte–in–cui–non | amministrazione–del–fondo–di–religione | ministro–di–grazia–e–giustizia | di–entrata–in–vigore–della |
| 16 | e–opposizione–agli–atti–esecutivi | repubblica–francese–una–ed–indivisibile | del–fondo–per–il–culto | del–decreto–del–presidente–della |
| 17 | oppure–confermano–principi–giurisprudenziali–pacifici | in–tutto–il–corso–della | e–dei–decreti–del–regno | da–un–punto–di–vista |
| 18 | nel–caso–in–cui–un | in–questo–stato–di–cose | atti–del–governo–a–f | la–questione–di–legittimità–costituzionale |
| 19 | nel–caso–in–cui–la | dal–punto–di–vista–della | raccolta–ufficiale–delle–leggi–e | per–la–prima–volta–nel |
| 20 | massimario–della–corte–di–cassazione | di–mano–in–mano–che | nella–gazzetta–ufficiale–del–regno | presidenza–del–consiglio–dei–ministri |
| 21 | data–di–entrata–in–vigore | del–primo–console–della–repubblica | ordiniamo–che–il–presente–decreto | in–tutto–o–in–parte |
| 22 | considerazione–del–fatto–che–il | ad–udire–la–parola–di | gli–uni–e–gli–altri | non–è–un–caso–che |
| 23 | con–la–conseguenza–che–il | la–quantità–di–calore–che | nella–raccolta–ufficiale–delle–leggi | a–che–fare–con–la |
| 24 | accertamento–del–giudice–di–merito | il–termine–di–giorni–quindici | nella–maggior–parte–dei–casi | di–stampare–nel–mese–di |
| 25 | sono–quelle–ufficiali–redatte–a | abilitato–ad–esercitare–la–professione | osservarlo–e–di–farlo–osservare | nel–momento–in–cui–si |
| 26 | sentenza–di–merito–che–aveva | gli–uni–che–gli–altri | iv–sezione–del–consiglio–di | ha–a–che–fare–con |
| 27 | rappresentanza–e–difesa–in–giudizio | a–poco–a–poco–si | munito–del–sigillo–dello–stato | finito–di–stampare–nel–mese |
| 28 | quelle–ufficiali–redatte–a–cura | del–fondo–di–religione | delle–poste–e–dei–telegrafi | anche–dal–punto–di–vista |
| 29 | pubblicate–sono–quelle–ufficiali–redatte | v–v–i–s–o | di–grazia–e–giustizia–e | dalla–data–di–entrata–in |
| 30 | nel–caso–in–cui–il | un–non–so–che–di | | |

**Table S20.** Top Italian 5-grams for arbitrary years.

| Rank | 1700 | 1800 | 1900 | 2000 |
|------|------|------|------|------|
| 1 | и | и | и | и |
| 2 | день | на | въ | в |
| 3 | на | не | на | на |
| 4 | въ | что | не | не |
| 5 | то | въ | что | что |
| 6 | а | по | по | по |
| 7 | въ | въ | его | к |
| 8 | не | его | для | как |
| 9 | города | а | какъ | в |
| 10 | го | для | а | из |
| 11 | пошли | то | къ | а |
| 12 | изъ | но | но | о |
| 13 | до | или | за | его |
| 14 | во | за | о | от |
| 15 | по | было | или | для |
| 16 | капитанъ | до | то | он |
| 17 | по | же | же | за |
| 18 | й | только | при | же |
| 19 | и | при | только | это |
| 20 | утру | о | было | я |
| 21 | пришли | къ | отъ | или |
| 22 | полки | они | изъ | но |
| 23 | того | того | это | то |
| 24 | къ | были | до | у |
| 25 | ь | бы | я | и |
| 26 | при | какъ | такъ | их |
| 27 | ночь | во | у | было |
| 28 | была | я | бы | только |
| 29 | что | у | время | до |
| 30 | также | отъ | въ | был |

Table S21. Top Russian 1-grams for arbitrary years.

| Rank | 1700 | 1800 | 1900 | 2000 |
|------|------|------|------|------|
| 1 | по–утру | и–не | и–въ | и–в |
| 2 | й–день | и–на | и–не | и–не |
| 3 | изъ–города | потому–что | и–на | в–том |
| 4 | также–и | и–по | не–только | а–также |
| 5 | во–вторникъ | во–время | такъ–какъ | не–только |
| 6 | го–года | и–въ | можетъ–быть | что–в |
| 7 | въ–неделю | не–было | но–и | но–и |
| 8 | того–же | и–въ | что–въ | так–и |
| 9 | однако–жъ | не–только | такъ–и | может–быть |
| 10 | на–день | что–на | не–было | о–том |
| 11 | въ–вечеру | для–того | и–по | и–на |
| 12 | была–изъ | что–они | потому–что | в–этом |
| 13 | шли–во | но–и | а–также | не–было |
| 14 | чинить–не | и–что | не–можетъ | что–он |
| 15 | февраля–въ | и–проч | какъ–и | и–его |
| 16 | сентября–день | что–не | во–время | как–и |
| 17 | пришли–и | а–не | что–онъ | из–них |
| 18 | пришелъ–и | что–въ | а–не | а–в |
| 19 | пргьхали–на | то–же | и–что | не–может |
| 20 | пошли–на | и–для | и–его | так–как |
| 21 | пошли–и | что–и | въ–томъ | таким–образом |
| 22 | пошелъ–на | на–него | о–томъ | а–не |
| 23 | потому–что | и–при | то–же | потому–что |
| 24 | полки–пришли | то–время | и–для | как–бы |
| 25 | по–городу | можно–было | у–него | не–в |
| 26 | перемення–лошадей | еще–не | въ–то | у–него |
| 27 | октября–день | но–не | еще–не | то–же |
| 28 | но–только | на–оборотъ | и–даже | в–его |
| 29 | не–могли | какъ–и | на–то | не–менее |
| 30 | не–дошедъ | не–можетъ | что–и | во–время |

Table S22. Top Russian 2-grams for arbitrary years.

| Rank | 1700 | 1800 | 1900 | 2000 |
|---|---|---|---|---|
| 1 | была–изъ–города | въ–то–время | то–же–время | то–же–время |
| 2 | чинить–не–могли | то–же–время | въ–то–время | по–отношению–к |
| 3 | по–городу–бить | на–другой–день | не–можетъ–быть | в–то–время |
| 4 | не–дошедъ–до | на–другой–день | въ–то–же | в–первую–очередь |
| 5 | хотя–они–отъ | мало–по–малу | бы–то–ни | не–может–быть |
| 6 | только–два–человека | бы–то–ни | то–ни–было | в–то–же |
| 7 | того–же–имени | и–того–ради | по–крайней–мере | по–крайней–мере |
| 8 | также–и–про | не–только–не | въ–это–время | в–то–же |
| 9 | также–и–день | то–ни–было | не–только–не | в–отличие–от |
| 10 | реку–на–другую | не–взирая–на | а–также–и | не–только–в |
| 11 | пушками–и–мортирами | въ–то–же | не–только–не | тем–не–менее |
| 12 | путь–и–шли | и–для–того | такъ–и–въ | тем–не–менее |
| 13 | пришли–къ–городу | въ–это–время | и–въ–то | так–и–в |
| 14 | пошли–на–море | не–иное–что | не–могутъ–быть | в–это–время |
| 15 | пошли–въ–путь | по–крайней–мере | хотя–и–не | так–и–в |
| 16 | потомъ–же–и | не–можетъ–быть | можно–было–бы | то–время–как |
| 17 | потому–что–онъ | и–во–время | равно–какъ–и | бы–то–ни |
| 18 | полки–перешли–на | въ–это–время | какъ–и–въ | как–и–в |
| 19 | поелѣ–того | преподобнаго–отца–нашего | но–и–въ | той–или–иной |
| 20 | подъ–его–командою | не–можно–было | а–равно–и | в–конце–концов |
| 21 | по–утру–рано | такъ–какъ–и | въ–это–время | а–также–в |
| 22 | по–а–что | а–на–оборотъ | къ–тому–же | к–тому–же |
| 23 | перешли–на–другую | хотя–и–не | въ–то–время | к–тому–же |
| 24 | переименовали–его–въ | по–тому–что | и–то–же | ко–и–в |
| 25 | они–только–въ | императора–петра–великаго | въ–первый–разъ | думы–федерального–собрания |
| 26 | ночь–на–день | и–при–томъ | на–этотъ–разъ | в–отличие–от |
| 27 | но–только–два | въ–то–время | одно–и–то | речь–идет–о |
| 28 | ничего–не–в | равно–какъ–и | и–такимъ–образомъ | можно–было–бы |
| 29 | никакого–вреда–его | не–что–иное | на–другой–день | в–том–же |
| 30 | не–токмо–что | на–что–на | по–этому–поводу | то–ни–было |

Table S23. Top Russian 3-grams for arbitrary years.

| Rank | 1800 | 1900 | 2000 |
|---|---|---|---|
| 1 | въ–то–же–время | въ–то–же–время | в–то–же–время |
| 2 | бы–то–ни–было | бы–то–ни–было | в–то–же–время |
| 3 | не–взирая–на–то | и–въ–то–же | в–то–время–как |
| 4 | что–на–что–на | одно–и–то–же | бы–то–ни–было |
| 5 | во–что–бы–то | одной–и–той–же | изменений–и–дополнений–в |
| 6 | какъ–бы–то–ни | одного–и–того–же | и–в–то–же |
| 7 | что–въ–то–время | въ–то–же–время | одно–и–то–же |
| 8 | что–бы–то–ни | что–бы–то–ни | в–той–или–иной |
| 9 | въ–то–же–время | какъ–бы–то–ни | совета–федерации–федерального–собрания |
| 10 | одно–и–то–же | во–что–бы–то | о–проекте–федерального–закона |
| 11 | и–въ–то–же | не–говоря–уже–о | так–же–как–и |
| 12 | что–въ–это–время | но–въ–то–же | сссp–по–обвинению–в |
| 13 | то–же–время–и | то–же–время–и | ни–в–чем–не |
| 14 | но–въ–то–же | въ–то–время–какъ | вквс–сссp–по–обвинению |
| 15 | не–задолго–до–того | и–то–же–время | федерального–собрания–российской–федерации |
| 16 | не–говоря–уже–о | отъ–времени–до–времени | один–и–тот–же |
| 17 | по–одному–или–по | и–въ–тоже–время | не–говоря–уже–о |
| 18 | по–два–и–по | и–не–можетъ–быть | одного–и–того–же |
| 19 | одному–или–по–два | какой–бы–то–ни | одной–и–той–же |
| 20 | одной–и–той–же | ни–за–что–не | что–бы–то–ни |
| 21 | не–взирая–ни–на | въ–одно–и–то | дополнений–в–федеральный–закон |
| 22 | не–было–ни–одного | что–въ–то–время | одни–и–те–же |
| 23 | на–что–на–что | не–можетъ–быть–и | во–что–бы–то |
| 24 | на–роды–и–виды | и–на–этотъ–разъ | государственной–думы–федерального–собрания |
| 25 | и–въ–то–время | не–говоря–уже–о | думы–федерального–собрания–российской |
| 26 | два–и–по–три | такъ–же–какъ–и | но–в–то–же |
| 27 | въ–то–же–время | одну–и–ту–же | и–дополнений–в–федеральный |
| 28 | въ–то–время–въ | одному–и–тому–же | и–тем–не–менее |
| 29 | узнать–было–не–можно | говоря–уже–о–томъ | постановление–государственной–думы–федерального |
| 30 | то–время–не–было | какихъ–бы–то–ни | то–же–время–в |

Table S24. Top Russian 4-grams for arbitrary years.

| Rank | 1800 | 1900 | 2000 |
|---|---|---|---|
| 1 | какъ–бы–то–ни–было | и–въ–то–же–время | и–в–то–же–время |
| 2 | во–что–бы–то–ни | какъ–бы–то–ни–было | вквс–ссср–по–обвинению–в |
| 3 | и–въ–то–же–время | во–что–бы–то–ни | во–что–бы–то–ни |
| 4 | но–въ–то–же–время | но–въ–то–же–время | государственной–думы–федерального–собрания–российской |
| 5 | по–одному–или–по–два | одно–и–то–же–время | думы–федерального–собрания–российской–федерации |
| 6 | по–два–и–по–три | какой–бы–то–ни–было | и–дополнений–в–федеральный–закон |
| 7 | кому–бы–то–ни–было | въ–одно–и–то–же | изменений–и–дополнений–в–федеральный |
| 8 | какого–бы–то–ни–было | въ–то–же–время–и | но–в–то–же–время |
| 9 | такъ–какъ–въ–это–время | какихъ–бы–то–ни–было | постановление–государственной–думы–федерального–собрания |
| 10 | одного–берега–озера–на–другой | какого–бы–то–ни–было | изменений–и–дополнений–в–закон |
| 11 | на–три–и–на–четыре | въ–одной–и–той–же | гд–постановление–государственной–думы–федерального |
| 12 | на–два–и–на–три | въ–то–же–время–не | в–то–же–время–в |
| 13 | которой–не–было–еще–и | какимъ–бы–то–ни–было | ш–гд–постановление–государственной–думы |
| 14 | какой–бы–то–ни–было | не–говоря–уже–о–томъ | в–то–время–как–в |
| 15 | во–что–бы–то–пи | но–въ–то–же–время | как–бы–то–ни–было |
| 16 | что–она–не–только–не | какъ–бы–то–ни–было | р–распоряжение–правительства–российской–федерации |
| 17 | что–ни–одна–птица–не | не–говоря–уже–о–томъ | в–той–или–иной–мере |
| 18 | что–въ–то–время–въ | не–можетъ–быть–и–речи | каких–бы–то–ни–было |
| 19 | потому–что–она–не–только | кого–бы–то–ни–было | какой–бы–то–ни–было |
| 20 | потому–что–въ–это–время | къ–одному–и–тому–же | а–в–дву–по–тому |
| 21 | потому–что–въ–то–время | на–одной–и–той–же | в–дву–по–тому–ж |
| 22 | по–три–и–по–четыре | на–одномъ–и–томъ–же | федерации–федерального–собрания–российской–федерации |
| 23 | одно–и–то–же–время | какое–бы–то–ни–было | совета–федерации–федерального–собрания–российской |
| 24 | но–не–взирая–на–то | какомъ–бы–то–ни–было | и–в–то–же–время |
| 25 | но–и–то–и–другое | не–одно–и–то–же | постановление–совета–федерации–федерального–собрания |
| 26 | но–и–между–ними–и | кому–бы–то–ни–было | но–в–то–же–время |
| 27 | ни–у–кого–не–брать | какими–бы–то–ни–было | в–одном–и–том–же |
| 28 | ни–для–кого–не–тайна | по–одному–и–тому–же | ни–в–чем–не–бывало |
| 29 | не–говоря–уже–о–томъ | въ–одномъ–и–томъ–же | ни–в–коей–мере–не |
| 30 | на–три–года–и–одиннадцать | къ–одной–и–той–же | как–ни–в–чем–не |

Table S25. Top Russian 5-grams for arbitrary years.

| Rank | 1700 | 1800 | 1900 | 2000 |
|------|------|------|------|------|
| 1 | de | de | de | de |
| 2 | que | que | la | la |
| 3 | y | y | que | y |
| 4 | la | la | y | que |
| 5 | el | en | el | en |
| 6 | en | el | en | el |
| 7 | los | los | los | a |
| 8 | con | se | á | los |
| 9 | á | á | del | del |
| 10 | se | las | se | se |
| 11 | no | del | las | las |
| 12 | por | por | por | por |
| 13 | del | con | no | un |
| 14 | las | no | con | con |
| 15 | lo | su | para | una |
| 16 | para | es | su | no |
| 17 | un | para | al | para |
| 18 | su | mas | un | su |
| 19 | le | lo | lo | es |
| 20 | es | un | una | al |
| 21 | a | al | es | como |
| 22 | al | una | como | lo |
| 23 | fe | como | a | o |
| 24 | mas | sus | ó | más |
| 25 | como | ó | sus | la |
| 26 | una | este | el | el |
| 27 | fu | esta | ha | sus |
| 28 | ha | le | i | en |
| 29 | me | si | más | sobre |
| 30 | tan | ha | la | entre |

**Table S26.** Top Spanish 1-grams for arbitrary years.

| Rank | **1700** | **1800** | **1900** | **2000** |
|---|---|---|---|---|
| 1  | de–la    | de–la    | de–la    | de–la    |
| 2  | de–los   | de–los   | de–los   | de–los   |
| 3  | en–la    | que–se   | en–el    | en–el    |
| 4  | en–el    | en–el    | que–se   | en–la    |
| 5  | lo–que   | de–las   | en–la    | a–la     |
| 6  | que–se   | en–la    | de–las   | de–las   |
| 7  | que–no   | á–la     | á–la     | que–se   |
| 8  | de–las   | de–su    | por–el   | a–los    |
| 9  | de–que   | lo–que   | que–el   | y–la     |
| 10 | con–el   | y–de     | á–los    | lo–que   |
| 11 | de–su    | á–los    | que–no   | de–un    |
| 12 | que–le   | que–no   | de–su    | y–el     |
| 13 | á–la     | que–el   | que–la   | en–los   |
| 14 | que–fe   | que–la   | lo–que   | que–la   |
| 15 | en–que   | en–los   | por–la   | que–el   |
| 16 | que–el   | y–que    | y–de     | de–su    |
| 17 | que–en   | y–la     | en–los   | de–una   |
| 18 | que–la   | en–que   | de–que   | y–de     |
| 19 | y–que    | que–en   | en–que   | por–el   |
| 20 | y–el     | de–que   | de–un    | con–el   |
| 21 | que–es   | y–el     | y–el     | por–la   |
| 22 | para–que | en–las   | en–las   | con–la   |
| 23 | por–la   | que–los  | y–la     | en–las   |
| 24 | con–la   | todos–los| con–el   | que–no   |
| 25 | á–los    | de–sus   | que–en   | a–las    |
| 26 | por–el   | no–se    | y–en     | de–que   |
| 27 | y–de     | y–en     | en–su    | en–su    |
| 28 | y–en     | los–que  | y–que    | en–un    |
| 29 | en–los   | con–la   | con–la   | de–sus   |
| 30 | en–las   | con–el   | de–una   | que–los  |

**Table S27.** Top Spanish 2-grams for arbitrary years.

| Rank | **1700** | **1800** | **1900** | **2000** |
|------|----------|----------|----------|----------|
| 1  | de–lo–que        | y–de–la          | de–la–república   | a–través–de       |
| 2  | uno–de–los       | en–que–se        | en–que–se         | uno–de–los        |
| 3  | a–que–respondió  | que–no–se        | de–la–ley         | y–de–la           |
| 4  | que–fe–le        | de–lo–que        | los–estados–unidos| a–partir–de       |
| 5  | de–que–se        | de–los–que       | y–de–la           | de–lo–que         |
| 6  | por–lo–que       | todo–lo–que      | que–no–se         | en–el–que         |
| 7  | de–los–que       | de–todos–los     | uno–de–los        | el–caso–de        |
| 8  | de–la–real       | lo–que–se        | de–que–se         | una–de–las        |
| 9  | en–que–se        | por–lo–que       | de–buenos–aires   | lo–que–se         |
| 10 | á–que–respondió  | por–medio–de     | presidente–de–la  | en–la–que         |
| 11 | rey–nueftro–señor| y–de–los         | de–lo–que         | de–la–república   |
| 12 | que–en–el        | de–que–se        | lo–que–se         | de–la–vida        |
| 13 | mil–ducados–de   | á–los–que        | que–se–le         | por–lo–que        |
| 14 | lo–que–se        | de–todas–las     | la–ley–de         | que–no–se         |
| 15 | en–lo–que        | que–en–el        | la–ciudad–de      | de–la–población   |
| 16 | el–duque–de      | de–la–tierra     | de–los–que        | de–la–sociedad    |
| 17 | que–se–le        | al–mismo–tiempo  | por–medio–de      | de–la–ciudad      |
| 18 | de–san–joseph    | uno–de–los       | de–la–misma       | y–a–la            |
| 19 | y–de–la          | se–ha–de         | de–los–estados    | en–que–se         |
| 20 | una–porcion–de   | de–los–hombres   | que–en–el         | parte–de–la       |
| 21 | se–ha–de         | que–se–ha        | de–la–ciudad      | a–pesar–de        |
| 22 | la–iglesia–de    | el–nombre–de     | de–todos–los      | la–necesidad–de   |
| 23 | veinte–y–quatro  | de–la–vida       | á–fin–de          | y–en–la           |
| 24 | que–en–la        | de–la–naturaleza | á–que–se          | y–en–el           |
| 25 | iglesia–de–san   | y–en–el          | una–de–las        | la–mayoría–de     |
| 26 | de–que–no        | la–mayor–parte   | y–en–el           | que–en–el         |
| 27 | que–es–el        | y–de–las         | y–de–los          | los–estados–unidos|
| 28 | lo–que–fe        | y–en–la          | que–se–ha         | de–que–el         |
| 29 | en–la–llama      | que–es–el        | la–comisión–de    | la–ciudad–de      |
| 30 | de–san–luis      | que–en–la        | y–en–la           | en–el–caso        |

**Table S28.** Top Spanish 3-grams for arbitrary years.

| Rank | 1700 | 1800 | 1900 | 2000 |
|---|---|---|---|---|
| 1 | la–iglesia–de–san | la–mayor–parte–de | presidente–de–la–república | a–través–de–la |
| 2 | iglesia–de–san–joseph | á–fin–de–que | de–los–estados–unidos | a–lo–largo–de |
| 3 | del–rey–nueftro–señor | por–medio–de–la | de–la–ley–de | en–el–caso–de |
| 4 | veinte–y–quatro–mil | mayor–parte–de–los | el–presidente–de–la | el–hecho–de–que |
| 5 | el–papel–azul–de | con–el–nombre–de | á–fin–de–que | el–punto–de–vista |
| 6 | á–un–fuego–fuerte | al–mismo–tiempo–que | en–el–caso–de | a–partir–de–la |
| 7 | y–quatro–mil–ducados | cada–uno–de–los | á–que–se–refiere | la–mayoría–de–los |
| 8 | en–la–forma–que | que–se–ha–de | la–mayor–parte–de | de–los–estados–unidos |
| 9 | un–baño–de–arena | de–lo–que–se | de–la–ciudad–de | con–el–fin–de |
| 10 | los–veinte–y–quatro | y–al–mismo–tiempo | lo–dispuesto–en–el | en–el–que–se |
| 11 | en–un–baño–de | de–todo–lo–que | cada–uno–de–los | cada–uno–de–los |
| 12 | en–la–llama–interior | no–es–mas–que | que–se–refiere–el | la–mayor–parte–de |
| 13 | por–la–via–seca | cada–una–de–las | de–que–se–trata | de–la–universidad–de |
| 14 | por–el–ácido–nitroso | es–lo–mismo–que | y–dése–al–registro | desde–el–punto–de |
| 15 | la–reyna–nueftra–señora | no–es–otra–cosa | de–la–provincia–de | en–la–ciudad–de |
| 16 | la–de–san–luis | que–se–han–de | dése–al–registro–nacional | la–década–de–los |
| 17 | en–un–crisol–de | del–mismo–modo–que | de–la–ley–de | de–la–ciudad–de |
| 18 | de–los–veinte–y | el–modo–con–que | en–la–ciudad–de | en–la–que–se |
| 19 | se–disuelve–en–el | de–los–que–se | dicho–francisco–de–villagra | en–el–caso–de |
| 20 | que–es–lo–que | es–uno–de–los | en–los–estados–unidos | la–medida–en–que |
| 21 | por–medio–del–soplete | para–que–no–se | con–el–objeto–de | en–el–sentido–de |
| 22 | mufla–de–un–horno | es–una–de–las | de–acuerdo–con–el | a–través–de–los |
| 23 | mil–ducados–de–ayuda | mayor–parte–de–las | el–caso–de–que | cada–una–de–los |
| 24 | mas–claro–que–el | circulacion–de–la–sangre | con–el–nombre–de | el–caso–de–la |
| 25 | la–mufla–de–un | en–el–año–de | de–la–comisión–de | a–la–hora–de |
| 26 | en–la–casa–del | todo–lo–que–se | días–del–mes–de | por–parte–de–los |
| 27 | el–hierro–y–la | con–el–fin–de | cada–una–de–las | de–la–ley–de |
| 28 | ducados–de–ayuda–de | en–el–caso–de | la–ciudad–de–santiago | a–lo–largo–del |
| 29 | de–un–horno–de | la–circulacion–de–la | de–mil–ochocientos–noventa | en–la–medida–en |
| 30 | de–la–real–hazienda | lo–que–se–ha | mil–ochocientos–noventa–y | en–el–marco–de |

**Table S29.** Top Spanish 4-grams for arbitrary years.

| Rank | 1700 | 1800 | 1900 | 2000 |
|---|---|---|---|---|
| 1 | veinte–y–quatro–mil–ducados | la–mayor–parte–de–los | el–presidente–de–la–república | desde–el–punto–de–vista |
| 2 | los–veinte–y–quatro–mil | la–circulacion–de–la–sangre | á–que–se–refiere–el | en–la–medida–en–que |
| 3 | en–un–baño–de–arena | la–mayor–parte–de–las | y–dése–al–registro–nacional | el–punto–de–vista–de |
| 4 | mufla–de–un–horno–de | no–es–otra–cosa–que | de–mil–ochocientos–noventa–y | a–lo–largo–de–la |
| 5 | mil–ducados–de–ayuda–de | que–la–mayor–parte–de | el–dicho–francisco–de–villagra | en–el–sentido–de–que |
| 6 | la–mufla–de–un–horno | por–lo–que–hace–á | los–señores–por–la–afirmativa | lo–que–se–refiere–a |
| 7 | se–disuelve–en–el–agua | por–la–gracia–de–dios | dirección–de–tierras–y–colonias | en–la–mayoría–de–los |
| 8 | de–los–veinte–y–quatro | de–la–mayor–parte–de | la–dirección–de–tierras–y | diario–oficial–de–la–federación |
| 9 | un–sabor–dulce–al–principio | no–es–mas–que–un | publíquese–y–dése–al–registro | la–segunda–mitad–del–siglo |
| 10 | mas–claro–que–el–del | de–mil–setecientos–treinta–y | la–mayor–parte–de–los | de–mil–novecientos–noventa–y |
| 11 | la–octava–parte–de–la | la–superficie–de–la–tierra | el–presidente–de–la–república | la–mayor–parte–de–los |
| 12 | la–mayor–parte–de–los | todas–las–partes–del–cuerpo | en–el–caso–de–que | de–la–década–de–los |
| 13 | es–semejante–á–la–que | no–es–mas–que–una | la–administración–general–del–estado | en–el–diario–oficial–de |
| 14 | en–las–paredes–del–vaso | en–la–mayor–parte–de | á–la–dirección–de–tierras | de–los–estados–unidos–mexicanos |
| 15 | en–la–mufla–de–un | en–el–tiempo–de–la | que–se–refiere–el–artículo | el–diario–oficial–de–la |
| 16 | en–la–iglesia–de–san | los–siglos–de–los–siglos | publíquese–y–dése–al–registro | en–el–campo–de–los |
| 17 | con–el–agua–de–cal | la–mayor–parte–de–la | lo–dispuesto–en–el–artículo | en–el–caso–de–la |
| 18 | que–quedó–sobre–el–filtro | lo–que–hace–á–la | la–ciudad–de–buenos–aires | desde–el–punto–de–vista |
| 19 | que–es–una–mina–de | la–parte–superior–de–la | de–estado–y–del–despacho | en–el–caso–de–los |
| 20 | poco–ó–nada–que–ha | por–lo–que–toca–á | lo–dispuesto–en–el–art | a–que–se–refiere–el |
| 21 | para–la–administracion–de–los | de–que–acabamos–de–hablar | estado–y–del–despacho–de | en–el–marco–de–la |
| 22 | las–paredes–del–vaso–una | y–esto–es–lo–que | de–la–ley–de–enjuiciamiento | américa–latina–y–el–caribe |
| 23 | la–sala–de–alcaldes–de | rey–de–la–gran–bretaña | los–estados–unidos–de–américa | de–cada–uno–de–los |
| 24 | en–la–iglesia–parroquial–de | á–cada–uno–de–los | de–los–estados–unidos–de | la–mayoría–de–los–casos |
| 25 | de–qualquier–calidad–que–sean | por–lo–que–toca–á | servicio–de–dios–y–de | en–lo–que–se–refiere |
| 26 | de–nuestra–señora–de–la | de–la–academia–de–las | desde–el–punto–de–vista | en–la–década–de–los |
| 27 | de–mil–setecientos–cincuenta–y | la–mayor–parte–de–una | la–mayor–parte–de–las | en–el–ámbito–de–la |
| 28 | de–la–iglesia–de–san | de–cada–una–de–las | han–de–uso–y–costumbre | a–lo–largo–de–los |
| 29 | de–esta–villa–de–madrid | por–lo–que–hace–á | de–casación–por–infracción–de | política–de–los–estados–unidos |
| 30 | a–los–reales–pies–de | como–se–puede–ver–en | de–la–cámara–de–diputados | constitución–política–de–los–estados |

**Table S30.** Top Spanish 5-grams for arbitrary years.


\end{document}